\documentclass[usenatbib]{mn3e} 
\usepackage{amsmath,amssymb}
\usepackage{tablefootnote}   
\usepackage{graphicx}         
\usepackage{gensymb}        
\usepackage{color}               
\usepackage{natbib}
\usepackage{times}
\usepackage{fixltx2e} 
\usepackage[flushleft]{threeparttable}
\usepackage{orcidlink}
  
\newcommand{\ramses}{{\small RAMSES}}
\newcommand{\ramsestext}{{\small RAMSES} }

\newcommand{\barolo}{$^{\scriptstyle 3\mathrm{D}}$B{\sc arolo} }
\newcommand{\barolop}{$^{\scriptstyle 3\mathrm{D}}$B{\sc arolo}}

\newcommand{\Msol}{\,{\rm M}_\odot} 
\newcommand{\Msolyr}{{\,\rm M}_\odot\,{\rm yr}^{-1}} 
\newcommand{\Mpc} {{\,\rm Mpc}}
\newcommand{\kpc} {{\,\rm kpc}}
\newcommand{\pc} {{\,\rm pc}} 
\newcommand{\K} {{\,\rm K}} 
\newcommand{\cc}{{\,\rm {cm^{-3}}}}

\newcommand{\kmsec}{{\,\rm {km\,s^{-1}} }}
\def\Gyr{\,{\rm Gyr}}
\def\Myr{\,{\rm Myr}}
\def\yr{\,{\rm yr}}
\def\um{\,{\rm \mu m}}

\newcommand{\ha}{\rm H$\alpha$ }
\newcommand{\Ha}{\rm H$\alpha$} 
\newcommand{\hb}{\rm H$\beta$ }

\newcommand{\sfr}{\rm SFR}
\newcommand{\Sfr}{{\rm SFR} }
\newcommand{\cloudy}{{\sc cloudy} }
\newcommand{\cloudyp}{{\sc cloudy}}
\newcommand{\lcars}{{\sc lcars} }
\newcommand{\lcarsp}{{\sc lcars}}
\newcommand{\hsim}{{\sc hsim} }
\newcommand{\hsimp}{{\sc hsim}}
\newcommand{\newhorizon}{{\sc NewHorizon} }
\newcommand{\newhorizonp}{{\sc NewHorizon}}
\newcommand{\eelt}{ELT }
\newcommand{\Eelt}{ELT}
\newcommand{\stromgren}{Str\"omgren sphere }
\newcommand{\stromgrenp}{Str\"omgren sphere}
\newcommand{\fpi}{${\rm (i)}$}
\newcommand{\fpii}{${\rm (ii)}$}
\newcommand{\fpiii}{${\rm (iii)}$}
\newcommand{\fpiv}{${\rm (iv)}$}
\newcommand{\fpv}{${\rm (v)}$}
\newcommand{\fpvi}{${\rm (vi)}$}
\newcommand{\fpvii}{${\rm (vii)}$}

\newcommand{\vlos}{v_{\rm LOS}}
\newcommand{\vlosma}{\langle\vlos\rangle} 
\newcommand{\vlosmi}{v_{\rm LOS,\,obs}}
\newcommand{\sigmasim}{\sigma_{\rm sim}}
\newcommand{\sigmaobs}{\sigma_{\rm obs}}
\newcommand{\sigmma}{\langle\sigmasim\rangle_{M}}
\newcommand{\vsigma}{v_{\theta}/\sigma}

\newcommand{\diff}{{\rm d}}
\newcommand{\ang}{{\,\rm\AA}}

\newcommand{\ergsec}{{\,\rm erg\,s^{-1}}}

\newcommand{\lunittl}{{\,\rm erg\,s^{-1}\,cm^{-2}\,\ang^{-1}\,arcsec^{-2}}}
\newcommand{\mas}{{\,\rm mas}}

\title[Determining Properties from Observations I] {On the Viability of Determining Galaxy Properties from Observations I: Star Formation Rates and Kinematics}\author[Kearn Grisdale et al.] 
{\parbox[t]{\textwidth}{Kearn Grisdale$^1$\thanks{kearn.grisdale@physics.ox.ac.uk}\orcidlink{0000-0003-0375-5997}, Laurence Hogan$^{1}$, Dimitra Rigopoulou$^{1}$, Niranjan Thatte$^{1}$, Miguel Pereira-Santaella$^{1,2}$, Julien Devriendt$^{1}$, Adrianne Slyz$^{1}$, Ismael Garc\'ia-Bernete$^{1}$, Yohan Dubois$^{3}$, Sukyoung K. Yi$^{4}$, Katarina Kraljic$^{5}$}\vspace*{6pt}\\
  	$^1$ Sub-department of Astrophysics, University of Oxford, Keble Road, Oxford OX1 3RH\\
	$^{2}$ Centro de Astrobiología (CSIC-INTA), Ctra. de Ajalvir, Km 4, 28850, Torrejón de Ardoz, Madrid, Spain\\
	$^{3}$ Institut d'Astrophysique de Paris, UMR 7095, CNRS, UPMC Univ. Paris VI, 98 bis boulevard Arago, 75014 Paris, France\\
	$^{4}$ Department of Astronomy, Yonsei University, 50 Yonsei-ro, Seodaemun-gu, Seoul 03722, Republic of Korea\\
	$^{5}$ Aix Marseille Univ, CNRS, CNES, LAM, Marseille, France\\}
\date{\today}
\begin{document}
\maketitle
\graphicspath{ {Figures/} }
\begin{abstract} 
We explore how observations relate to the physical properties of the emitting galaxies by post-processing a pair of merging $z\sim2$ galaxies from the cosmological, hydrodynamical simulation \newhorizon using \lcars (Light from Cloudy Added to RAMSES) to encode the physical properties of the simulated galaxy into \ha emission line. By carrying out mock observations and analysis on these data cubes we ascertain which physical properties of the galaxy will be recoverable with the HARMONI spectrograph on the European Extremely Large Telescope (\Eelt). We are able to estimate the galaxy's star formation rate and dynamical mass to a reasonable degree of accuracy, with values within a factor of $1.81$ and $1.38$ of the true value. The kinematic structure of the galaxy is also recovered in mock observations. Furthermore, we are able to recover radial profiles of the velocity dispersion and are therefore able to calculate how the dynamical ratio varies as a function of distance from the galaxy centre. Finally, we show that when calculated on galaxy scales the dynamical ratio does not always provide a reliable measure of a galaxy's stability against gravity or act as an indicator of a minor merger.
\end{abstract}


\section{Introduction}
\label{sect:intro}

Observations are the primary means by which humanity explores the Universe. Over the next several years new facilities such as the (European) Extremely Large Telescope (\Eelt), with its $39 {\rm\,m}$ diameter primary mirror, will become operational and present new observational opportunities. Such telescopes will allow higher resolution and deeper observations than current facilities \citep{Tamai:2016aa,Cirasuolo:2020aa}. This will be pivotal for the study of galaxy formation and evolution, particularly at high redshifts ($z\gtrsim1.5$).
It is therefore vitally important to understand how these observations translate back into the intrinsic properties of galaxies, giant molecular clouds (GMCs), the interstellar medium (ISM) and star formation. 

Through multiwavelength observations researchers have been able to determine the Star Formation Rate ($\sfr$) of high redshift galaxies. Such observations have resulted in the measurement of a correlation between \sfr\, and stellar mass ($M_{\star}$) of a galaxy which has since become known as the Main Sequence (MS) of star forming galaxies \citep[e.g.][]{Daddi:2007aa, Noeske:2007aa}. Star forming galaxies can be placed into one of two populations: normal star forming or starbursts. Normal star forming galaxies lie within the scatter of the MS correlation ($\sim0.3{\rm\,dex}$) whereas starbursts are defined as having a $\sfr$ at least four times greater than expected from the MS correlation  \citep{Elbaz:2011aa, Schreiber:2015aa}. Furthermore, it has also been determined that the cosmic star formation rate density (SFRD) evolves with $z$, having peaked at $z\sim1.9$, before declining by an order of magnitude to the current epoch \citep{Madau:2014aa}. The normalisation of the MS correlation increases with redshift \citep[][]{Speagle:2014aa} in line with the increase in the cosmic SFRD, which means galaxies at high-$z$ have higher $\sfr$s than those at the current epoch but are not be considered starburst galaxies.
For example, at $z\sim2$ galaxies on the MS tend to have a \sfr\, of $\sim20\times$ that of an MS galaxy at $z=0$. 

A relationship between the gas surface density ($\Sigma_{\rm g}$) and \sfr\, surface density ($\dot\Sigma_{\star}$) of galaxies has also been found, often taking the form of a so-called Kennicutt-Schmidtt law: $\dot\Sigma_{\star}\propto \Sigma_{\rm g}^{n}$, with $n\sim1.5$ \citep[][]{Schmidt:1959aa,Kennicutt:1998aa}. The scalings of \sfr\, with $z$ and $\Sigma_{\rm g}$ support observations showing that galaxies at high $z$ have more gas than galaxies in the local Universe. 

Measuring the kinematics of a galaxy can provide key insights into the processes guiding its evolution, such as the source of dynamical support \citep[][]{Puech:2007aa,Epinat:2009aa} or determining if a galaxy is undergoing a merger. For example, both the K-band Multi Object Spectrograph (KMOS) Redshift One Spectroscopic Survey (KROSS) and KMOS$^{\rm 3D}$ surveys \citep[see][respectively and references within]{Stott:2016aa, Wisnioski:2019aa} used Integral Field Spectroscopy (IFS) observations of \ha emissions line to calculate rotational velocities and velocity dispersions for a number of galaxies at $0.6<z<2.7$. These studies were able to determine whether or not a given galaxy in their sample was rotationally dominated. Such determinations can be made by comparing the rotational velocity to the velocity dispersion \citep[commonly known as $\vsigma$ and first introduced by][]{Binney:1978aa}. Alternatively, it is possible to use stability criteria such as the classic Toomre ``$Q$'' \citep[][]{Toomre:1964aa} or the more recent Romeo dispersion relation \citep[][]{Romeo:2010aa}.

Recently, \cite{Hogan:2021aa}, hence forth H21, performed observations, using the KMOS on the Very Large Telescope (VLT), of a sample of luminous Infra-Red (IR) galaxies, at $2 < z < 2.5$. The goal of this work was to establish whether $z\sim2$ luminous IR galaxies are isolated or interacting discs. H21 established that a significant fraction, $\sim40\%$, appear to be isolated galaxies based on their properties, such as $\vsigma$, their position on the MS, and their levels of dust obscured star formation when compared to local starburst galaxies. H21 determined that the huge \sfr\, seen in IR luminous galaxies at cosmic noon can be powered by steady state mechanisms and do not require stochastic events, such as the gas rich major mergers that power local ultra luminous infrared galaxies.

All of these observations require some method for converting the photons received from a galaxy into physical measurements of its properties, e.g. multiplying the integral of an emission line by a conversion factor to get a \sfr\, \citep[see][]{Kennicutt:1994aa}. Cosmological simulations can provide useful insight into observations by running their suites,  each wit different physical models, and comparing the results to observations. For example, \cite{Grisdale:2017aa} compared the power spectrum calculated from H{\small I} observations \citep[see][]{Walter:2008aa} with the spectrum calculated from simulations. In this study it was found stellar feedback processes were needed to recreate the observed power law slopes. 

In the last several years it is becoming increasingly computationally viable to run cosmological simulations which include radiative transfer \citep[RT, for example see][]{Rosdahl:2013aa,Kannan:2019aa}. Running such simulations, particularly on larger volumes takes a significant number of CPU hours. The other drawback to these simulations is they tend to have a severely limited spectral resolution, i.e. only having $N$ wavelength bins, where $N$ is normally between 1 and 10 \citep[e.g.][used 3 bins which covered wavelengths between $0$ and $910\ang$]{Rosdahl:2018aa}. Thus these kinds of RT-cosmological simulations can provide insight into the number of photons in a given wavelength range however they are unable to provide information about the shape and position of emission features. This makes creating line of sight velocity or velocity dispersion maps more difficult. Additionally, all the parameters of an RT simulation, such as the assumed shape of the star's Spectral Energy Distribution (SED), is set before run time and exploring how the choice of SED would affect the observation of the simulated galaxy requires re-running the simulation, adding to the computational cost. For a more holistic view RT simulations we direct the read to \cite{Iliev:2009aa}.

One way to gain higher spectral resolution and be able to test how changing the parameters affect observability is to employ a post-processing pipeline to ``paint'' photons into the simulation after runtime. In \cite{Grisdale:2021ab} we employed such a pipeline to explore the likelihood of detecting Population III stars with the \Eelt. We were able to show that if such stars have top heavy Initial Mass Functions (IMF) detection would be possible. 
 
 The High Angular Resolution Monolithic Optical and Near-infrared Integral field spectrograph (HARMONI) will be the work-horse spectroscopic instrument on \Eelt, providing spectra from $0.47$ to $2.45\um$ \citep[][]{Thatte:2014aa}. This relatively large wavelength range and its high spatial resolution coupled with the large primary mirror of the \Eelt, makes it even more important to ensure we are able to accurately interpret observations and decode the properties of the emitting object. It is here that processing simulations with pipelines, such as the one described used in \cite{Grisdale:2021ab}, can be extremely powerful. 
 
 The primary goal of this work is to explore how the physical properties of a galaxy can be \emph{accurately} recovered from (simulated) observations of emission lines produced in the galaxy. To that end we make use of the hydrodynamical simulation post-processing pipeline \lcars (Light from Cloudy Added to RAMSES) to convert properties, such as the star formation rate and gas kinematics, into photons which can then be ``observed'' using the HARMONI simulator (\hsimp) before being analysed. In Section~\ref{sect:meth} we outline \lcars as well as the simulation suite used in this work. We present our galaxy selection criteria, chosen galaxy and its intrinsic properties in Section~\ref{sect:galaxydiscrip}, while Section~\ref{sect:resultsi} explores the measured properties of this galaxy after being observed. We discuss implications of spatial resolution, whether photons are a good tracer for gas properties and the meaning of $\vsigma$ in the new high spatial resolution observation paradigm in Section~\ref{sect:discusion}. Section~\ref{sect:conclusion} presents a summary of this work and its conclusions.

\section{Method}
\label{sect:meth}

\subsection{Simulation Data Set: \newhorizon Overview}
\label{meth:nh}
In this work we have selected a simulated galaxy (see \S\ref{sect:galaxydiscrip} for details of this galaxy) from the \newhorizon simulation. \newhorizon is a hydrodynamical, cosmological simulation run using the hydro+$N$-body, Adaptive Mesh Refinement (AMR) code {\ramses} \citep{Teyssier:2002aa} targeting a linear spatial resolution for the smallest cell of $\Delta x\sim35\pc$. New levels of refinement are unlocked as the simulation progresses to account for cosmological expansion. Here we give a very brief overview of the simulation but direct the reader to \cite{Dubois:2021aa} for complete details \citep[see also][]{Park:2019aa,Grisdale:2021ab}. 

\newhorizon is a re-simulation of a spherical region with a radius of $10\Mpc$ (comoving) which has been extracted from the Horizon-AGN simulation \citep{Dubois:2014aa,Kaviraj:2017aa}. The simulation includes physical processes such as star formation, stellar feedback from star particles and Active Galactic Nuclei (AGN) feedback from black hole particles. However it does not include explicit radiative transfer. Star formation occurs on a cell by cell basis when a cell's gas number density is $\geq10{\rm\, cm}^{-3}$ and temperature $<2\times10^{4}{\rm\,K}$, following a Schmidt law \citep{Schmidt:1959aa}. The star formation efficiency per free-fall time is determined by the local gravo-turbulent conditions of the ISM \citep[][]{Kimm:2017aa}. Each star particle has a mass of $M_{\star}\geq10^{4}\Msol$ and is assumed to represent a population of stars following a Chabrier IMF \citep[][]{Chabrier:2005aa} with lower and upper mass cutoffs of $0.1\Msol$ and $150\Msol$ respectively. These particles in turn inject momentum ($3\cdot10^{49}\,{\rm erg}\Msol^{-1}$) back into the ISM via supernovae explosions occurring $5\Myr$ after each particle forms \citep[][]{Kimm:2014aa}. 

The simulation produces black holes (BH) particles in cells where both gas and stellar densities are $>10{\rm\, cm}^{-3}$. These particles are initialised with a seed mass of $10^{4}\Msol$. The mass of the particles can increase through Bondi--Hoyle--Lyttleton accretion and through BH coalescence\citep[capped at the Eddington limit, ][]{Bondi:1944aa,Hoyle:1939aa}. BH particles inject feedback into their environment via one of two modes, radio or quasar, set by the accretion rate of gas onto the BH relative to Eddington \citep[][]{Dubois:2012aa}. Dark matter (DM) particles are also included in the simulation, modelled as collision particles with a mass of $10^{6}\Msol$.

\subsection{Light from Cloudy Added to \ramsestext (\lcarsp)}
\label{meth:lcars}
In \cite{Grisdale:2021ab}, henceforth G21, we introduced a post processing pipeline that added photons into \ramsestext simulations such as \newhorizon. This pipeline took the properties of star particles and gas within the simulation and combined these with a large number of radiative transfer simulations using the microphysics code \cloudy \citep[see][for full details on \cloudyp]{Ferland:2017aa} to determine the shape and magnitude of an emission line along each line of sight. In this work we make use of an improved version of this pipeline. We refer to this pipeline as ``Light from Cloudy Added to \ramses'' or \lcarsp. 

\subsubsection{Single ``Super'' Star Particle in Cell Approach}
\label{lcars:ssp}
\begin{figure}
	\begin{center}
		\includegraphics[width=0.45\textwidth]{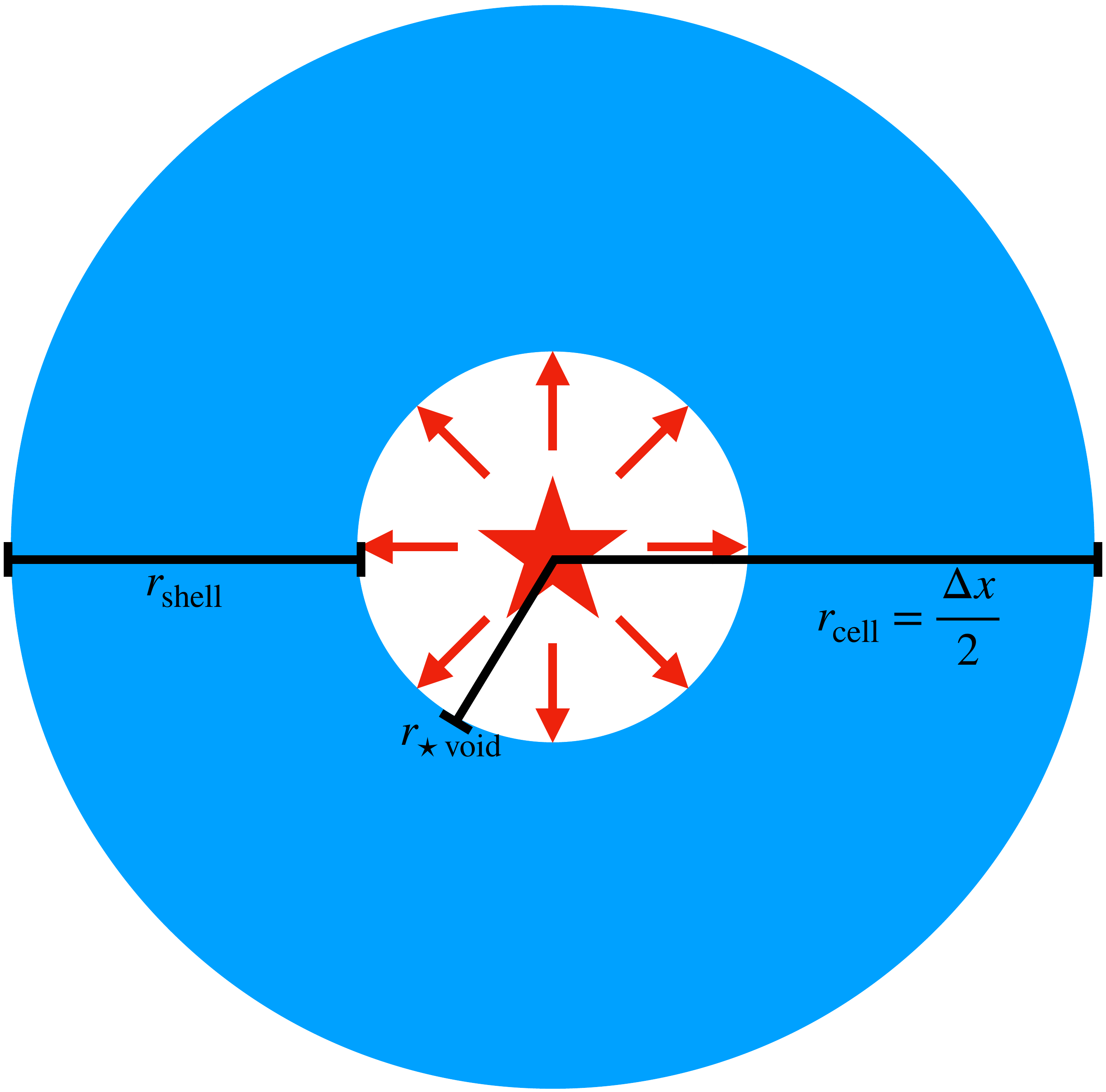} 
		\caption{2D depiction of a SSP in a cell. The red star and red arrows represent the SSP and the radiation/wind that has evacuated the void (white) region of gas. The cell's gas is shown by the blue shell surrounding the SSP. The black lines indicate the scale of each part of the diagram. }
	\label{fig:cellcartoon}
	\end{center}
\end{figure}
The previous version of \lcars (see G21) required running a grid of \cloudy simulations that covered all possible combinations of cell properties and stellar properties. Each star particle was then matched to the \cloudy simulation that best matched it and its host cell. For cells with multiple particles, each particle was treated as if it was the \emph{only} one within its host cell when matching to \cloudyp. The final spectrum of such a cell was created from the summation of the matched \cloudy outputs of each individual particle within the cell.
In this updated version of \lcars we assume that all star particles within a given cell are located at its centre, in one massive ``super'' star particle (SSP). 
A unique SED is calculated for each SSP by summing the individual SED of each of its constituent particles. We assume that the SSP is situated at the centre of cell and has removed all gas, via wind and radiation, within the radius, $r_{\star\,\rm void}$, where 
\begin{equation}
	r_{\star\,\rm void} = \left(\frac{3M_{\rm SSP}}{4\pi\rho_{0,\star}}\right)^{1/3},
	\label{eq:rviod}
\end{equation}
or $\Delta x \times 0.45$, whichever is smaller. Here $M_{\rm SSP}$ is the mass of the SSP while $\rho_{0,\star}=1500\Msol\pc^{-3}$ is a constant stellar density.\footnote{The value $\rho_{0,\star}$ is derived by dividing the average mass of a star particle by the size of the void region ($r_{\star\,\rm void}=2\pc$) used in G21.} 
To maintain mass conservation the gas density of the cell must be adjusted to account for the void region surrounding the SSP. This is achieved by evenly distributing the gas mass throughout a spherical shell with a thickness $r_{\rm shell} =\left(\Delta x/2\right)-r_{\star\,\rm void}$,  see Fig.~\ref{fig:cellcartoon}. This modified gas density ($n_{\rm cloud}$) is then used as an input parameter for \cloudyp.  


\subsubsection{Constructing Spectra and Data Cubes}
\label{lcars:spec}
For each cell with an SSP a unique \cloudy simulation is run. Each of these \cloudy simulations are run assuming spherical shell of gas surrounding a single primary radiation source (the SSP), as shown in Fig.~\ref{fig:cellcartoon}. The input parameters for \cloudyp, listed in Table~\ref{table:cloudy}, are set individually for each cell. In addition to the SSP radiation source our \cloudy simulations also include diffuse light from neighbouring cells (see \S\ref{lcars:diffuse}). A background radiation field which mimics the observed cosmic radio to X-ray background, with contributions from the CMB, is also included. This is assumed to be a black body with a temperature of $T_{\rm CMB}=2.725(1+z){\rm K}$.

\ramsestext calculates the gas temperature ($T_{g}$) for each gas cell within the simulation volume at run time. As a result of the supernovae feedback prescriptions included in \newhorizonp, there are cells with $T_{g}$ in excess of $10^{4}\K$, with some reaching a few $10^{8}\K$ \citep[we refer the reader to][for full details on the heating and cooling mechanisms within the simulations]{Dubois:2021aa}. We note that only $\sim0.003\%$, by mass, of G1's gas has $T_{g}\geq10^{8}\K$. \cloudy is unaware of when or where supernovae have occurred and therefore will never predict temperatures $\gtrsim2\cdot10^{4}\K$. As a result \cloudy will not correctly predict the ionisation state of the gas in such hot cells. To address this we employ a temperature threshold: when running \cloudy models for cells with $T_{g}>2\cdot10^{4}\K$ we enforce a constant temperature equal to $T_{g}$. For the cells below this threshold we allow \cloudy to determine the temperature. The temperature calculated by \cloudy is \emph{only} used during \cloudyp's radiative transfer calculations and not for determining line width.

The strength of each emission line ($I_{\rm 0}$) and the continuum at $r_{\rm shell}$, as calculated by \cloudyp, are then used to construct a spectrum for each cell. First we assume the emission line is given by  
\begin{equation}
	I(\lambda) = I_{\rm norm}e^{-\frac{(\lambda-\lambda_{\rm c})^{2}}{2\sigmasim^{2}}},
	\label{eq:gaus}
\end{equation}
where $I_{\rm norm}$ is a normalisation constant ensuring that $\int I(\lambda)\diff\lambda=I_{\rm 0}$, $\lambda_{\rm c}$ is the wavelength of the line in that cell and $\sigmasim$ sets the width of the line. $\lambda_{\rm c}$ is related to the emission line's rest wavelength ($\lambda_{\rm e}$) by 
\begin{equation}
	\lambda_{\rm c}=\lambda_{\rm e}(\frac{\vlos}{c}+1),
	\label{eq:lshift}
\end{equation}
here $\vlos$ is the bulk velocity of the gas within the cell along the line of sight (with respect to the observer). The width of the emission line is set by a combination of both the thermal motions of the gas and velocity dispersion of the gas across the cell, i.e. 
\begin{equation}
	\sigmasim =\sqrt{\sigma_{\rm g,therm}^{2}+\sigma_{\rm g,disp}^{2}},
	\label{eq:gasbroad0}
\end{equation}
where
\begin{equation}
	\sigma_{\rm g,therm}= \lambda_{\rm c}\sqrt{\frac{k_{B}T_{g}}{m_{a} c^{2}}},
	\label{eq:gasbroad1}
\end{equation}
and
\begin{equation}
	\sigma_{\rm g,disp}=\sqrt{(\sigma_{g,x}^{2}+\sigma_{g,y}^{2}+\sigma_{g,z}^{2})/3}.
	\label{eq:gasbroad2}
\end{equation}
$T_{g}$ is the gas temperature in the cell provided by \newhorizonp, $m_{a}$ is the mass of the element emitting the line, $c$ is the speed of light, $k_{B}$ is the Boltzmann constant, 
$\sigma_{g,x},\,\sigma_{g,y}$ and $\sigma_{g,z}$ are the gas velocity dispersions of the gas in cell along the three spatial axes of the simulation. The Full Width Half Maximum (FWHM) of the emission line is related to $\sigmasim$ by FWHM$\,= 2\sqrt{2\ln2}\sigmasim$. 
The emission line is then added to the continuum\footnote{Before addition the continuum is Doppler shifted to account for $\vlos$.} calculated by \cloudyp. 

Extinction is then applied to the spectrum through 
\begin{equation}
	I_{\rm E}(\lambda)=I(\lambda)10^{\frac{A_{V}E(\lambda)}  {-2.5}},
\end{equation}
where $I(\lambda)$ is the spectrum of a cell before extinction, $I_{\rm E}(\lambda)$ is the spectrum after extinction is applied and $E(\lambda)$ is the dust extinction curve found by \cite{Fitzpatrick:1999aa}. $A_{V}$ is unique to each cell and given by 
\begin{equation}
	A_{V}=1.086\frac{3f_{\rm d}\Sigma_{\rm Z}Q_{\lambda}}{4\rho_{\rm d}\sqrt{a_{1}a_{2}}},
	\label{eq:av}
\end{equation}
where $\Sigma _{\rm Z}$ is the column density of metals, $\rho_{\rm d}=3{\rm \,g\,cm^{-3}}$ is the typical density of a dust particle, $Q_{\lambda}=1.5$ is a constant extinction coefficient,  $f_{\rm d}=0.01$ is the fraction of gas-phase metals locked up in dust, $a_{1}=0.005{\rm \,\mu m}$ is the smallest size of a dust grain and $a_{2}=1{\rm \,\mu m}$ is the largest size of a dust grain \citep[see][for full details of the method and discussion on choice of values]{Richardson:2020aa}. 
$\Sigma _{\rm Z}$ is unique to each cell as it depends on the mass of metals along the line of sight between a cell and the observer. 

$f_{\rm d}$ is a free parameter which can be varied to improve the accuracy of this dust extinction model. With $f_{\rm d}=0.01$ we find our target galaxy (see \S\ref{sect:galaxydiscrip}) has an $A_{V}$ of $3.44\,{\rm mag}$ when calculated using a single, galaxy wide spectrum with the methods outlined in \S\ref{res:lcars:sfr}. Calculating an $A_{V}$ map of the galaxy directly from the simulation we find that most extinction line of sight through the galaxy to be $A_{V}\sim22\,{\rm mag}$\footnote{When calculating the $A_{V}$ map we take the value of $A_{V}$ straight from Eq.~\ref{eq:av}. Calculations of extinction from the \lcars SSC use the spectrum of the entire galaxy which results in information about extinction as function of depth being lost} and is located in the densest part of a spiral arm.  Increasing $f_{\rm d}$ by a factor of ten leads to values of $7.46\,{\rm mag}$ and $\sim224\,{\rm mag}$, respectively. From observations it could be argued that a value of $0.2$ is more reasonable \citep{Peeples:2014aa}, however we find $f_{\rm d}=0.01$ to produce a more realistic $A_{\rm V}$ value for a $z=2$ galaxy \citep[for example see][]{Kahre:2018aa}. 

Extinction due to the gas within the current cell is not included as this is applied by \cloudyp. As in G21, we note that the above extinction model is rather simple and does not account for every process that is able to reduce the strength of an emission line or spectra. So that the impact of extinction can be explored a second cube of each line is produces which does \emph{not} include the extinction. 

To produce a Spatial-Spectral Cube (SSC), we sum the spectra from all cells along sight line. Finally the wavelengths of the SSC are redshifted to match the value of $z$ of the simulation at the time of observation, and the SSC is divided by $4\pi D^{2}_{\rm L}\delta a^{2}$ to account for the luminosity distance ($D_{\rm L}$) and the size of the cell in arcseconds ($\delta a$). The produced SSC has units of $\lunittl$.

\subsubsection{Constructing Spectra for Cells Without Stars}
\label{lcars:nostar}
\begin{table}

	\parbox{0.4\textwidth}{	
		\caption{\cloudy Parameters}
		\begin{tabular}[h]{l l }		
			\hline \hline
			Parameter	           & Definition                                                           \\
			\hline
			SSP's SED              & Sum of SEDs for star particles within the SSP                        \\
			$L_{\rm SED,tot}$      & Integral of the SSP's SED, i.e. SSP's total luminosity               \\
			$r_{\star\,\rm void}$  & Radius of the gas empty region surrounding SSP                       \\
			$n_{\rm cloud}$        & Cell's gas number density (modified for assumed geometry)            \\
			$Z$                    & Metallicity of the gas in cell                                       \\
			$z$                    & Redshift of the simulation                                           \\
			$r_{\rm shell}$        & Thickness of gas spherical shell, $(\Delta x/2)-r_{\star\,\rm void}$ \\ 
			$T_{g}$                & Temperature of the gas (used in certain conditions)                  \\                                         
			\hline\hline
		\end{tabular}\\
		\label{table:cloudy}
		}	
\end{table}

In \S\ref{lcars:ssp} and \ref{lcars:spec} we limited the discussion to cells that contains star particles. However, in any sufficiently large volume extracted from \newhorizon there are cells which do not contain any star particles or diffuse emissions from neighbouring cells (see \S\ref{lcars:diffuse}). For these cells we run a grid of \cloudy simulations varying $n_{\rm cloud}$ and $Z$ (see Table~\ref{table:cloudy}) to cover the range of possible values found in these cells. 
Given the lack of photons to provide heating and that the majority of these cells are found within the hot Circumgalactic Medium (CGM) or hot (supernova heated) bubbles, to ensure the best match between \cloudy and \newhorizonp, we opt to also include $T_{g}$ as a third parameter in the grid of \cloudy simulations. 
These \cloudy simulations still assume a spherical shell geometry but with $r_{\star\,\rm void}=0$ and $r_{\rm shell}=\Delta x/2$. Furthermore, the incident radiation field used in these simulations is just the background radiation field described in \S\ref{lcars:spec}. The spectrum in these cells is constructed and added to the SSC in an identical manner to those containing a SSP.

\subsubsection{Diffuse Light From Neighbouring Cells}
\label{lcars:diffuse}
We use the radius of the hydrogen \stromgren \citep{Stromgren:1939aa}, i.e.
\begin{equation}
	R_{\rm SS,H} = \left(\frac{3Q_{H}}{4\pi n_{H} n_{\rm e} \alpha_{{\rm B},H}}\right)^{1/3},
	\label{eq:ssphere}
\end{equation}
 as a measure of the fraction of light from a given SSP that has escaped its host cell. Here $Q_{\rm H}$ is the number of photons per second with sufficient energy to ionise hydrogen, $n_{H}$ is the number density of hydrogen, $\alpha_{{\rm B},H}$ is the hydrogen recombination rate and $n_{\rm e}$ is the number density of electrons. Both $n_{H}$ and $n_{\rm e}$ are taken from the simulation and not account for $r_{\star\,\rm void}$. If $R_{\rm SS,H}\leq0.5\Delta x$ we assume that all ionising photons interact with the gas in the host cell of the SSP. 

\begin{figure*}
	\begin{center}
		\includegraphics[width=1.0\textwidth]{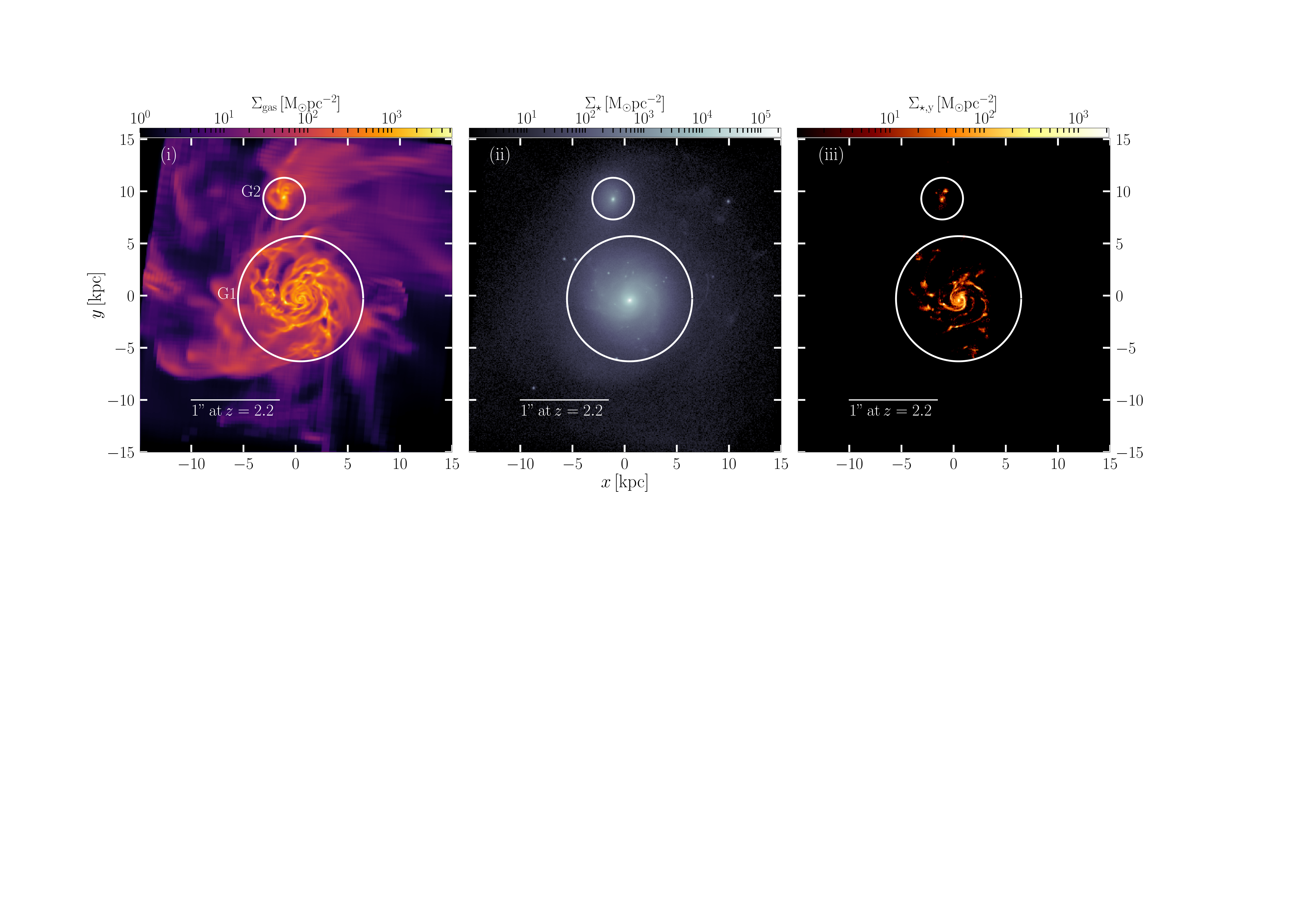} 
		\caption{Surface density maps for G1 and G2, calculated directly from the simulations. The large white circle indicates the primary galaxy (G1), while the small white circle shows the position of the merging companion galaxy (G2). These two circles have radii of $6\kpc$ and $2\kpc$ respectively. Panels \fpi-\fpiii\, show: the gas surface density $(\Sigma_{\rm gas})$, stellar surface density $(\Sigma_{\rm \star})$ and the young stellar surface (i.e. stars with ages $\leq10\Myr$) density $(\Sigma_{\rm \star,y})$. }
	\label{fig:smaps}
	\end{center}
\end{figure*}
In the case of the \newhorizon galaxy described in \S\ref{sect:galaxydiscrip}, we find that $\sim18\%$ of SSPs have a $R_{\rm SS,H}>0.5\Delta x$, \lcars therefore needs to be able to account for photons from one cell being deposited into the neighbouring cells. For each SSP with $R_{\rm SS,H}>0.5\Delta x$ this is achieved by:
\begin{enumerate}
	\item Identifying and tagging all cells which intersect with the \stromgren of the emitting SSP as ``contaminated''.
	\item The fraction ($f_{\rm SS}$) of the \stromgrenp's volume contained within each contaminated cell is calculated.
	\item The SED of the emitting SSP is multiplied by $f_{\rm SS}$ and added to the contaminated cell. 
	\item The SED added to each contaminated cell is subtracted from the SED of the emitting SSP.
	\item Results from (iii) and (iv) are stored, while the original SED of both the contaminated and emitting cells remain unchanged.
	\item Once all SSP's with $R_{\rm SS,H}>0.5\Delta x$  are identified and the contamination of all cells is calculated, the SED of each cell is updated to account for both contamination and leaked light. 
\end{enumerate}
Steps iv-vi ensure that the total energy of each SED is conserved during the diffusion calculation. Any cell that does not contain a SSP but is contaminated by a neighbouring cell is flagged and treated as if it does contain a SSP with the SED set purely by the contamination. More simply: the shape of the SED of such a contaminated cell is the sum of all SEDs contaminating that cell.

This method to account for diffuse light is only a first order approximation since it does not account for differences in the gas density of a contaminated cell compared to the emitting cell and the resulting change in the \stromgrenp. Furthermore, we use the hydrogen \stromgren as our marker for photons escaping the host cell, which neglects the fact that different elements will be ionised out to a different radius from the ionising source, i.e. each element will have different sized {\stromgrenp}s. In this work we are focused solely on hydrogen emissions lines making the hydrogen \stromgren an excellent proxy.

\subsubsection{Choice of SED}
\label{lcars:sed}
In order for the SED of a SSP to be created we required SEDs for each of its constituent star particles. To that end we employ {\sc starburst99} \citep[see][for details]{Starburst99} to generate a range of different SEDs which cover the metallicity  and age of the star particles found in \newhorizonp. 

We create SEDs with {\sc starburst99} by assuming a stellar population with an initial mass of $10^{5}\Msol$ and  the same Chabrier IMF used by \newhorizon at runtime. By allowing the stellar population to evolve following the Geneva Standard evolution tracks, for a given metallicity, {\sc starburst99} produces SEDs for a given population at ages between $10^{4}$ and $10^{9.7}\yr$ .  We generate one set of SEDs for each of the five different discrete metallicities offered by {\sc starburst99}. 

At runtime \lcars matches each star particle to the SED closest in age and metallically. As star particles formed in \newhorizon have a variety of birth masses ($M_{\star,\rm\,birth}$) it is necessary to scale the magnitude of the SED by $M_{\star,\rm\,birth}/10^{5}$. This produces the same result as running {\sc starburst99} with individual values of $M_{\star,\rm\,birth}$.

\begin{table}

		\parbox{0.4\textwidth}{	
			\caption{{\sc Summary of Galaxy Properties}}
			\begin{tabular}[h]{l r r r }		
				\hline \hline
				Property                               & G1      & G2   & G2/G1    \\
				                                       &         &      &          \\
				\hline
				$\log_{10}(M_{\rm gas}\,[\Msol])^{1}$  &   10.2  &  9.3 & $12.6\%$ \\  
				$\log_{10}(M_{\star}\,[\Msol])^{2}$    &   10.7  &  9.4 & $ 5.0\%$ \\  
				$\log_{10}(M_{\rm B}\,[\Msol])^{3}$    &   10.9  &  9.6 & $ 5.0\%$ \\  
				$\log_{10}(M_{\rm DM}\,[\Msol])^{4}$   &   10.7  &  9.3 & $ 4.0\%$ \\  
				$\log_{10}(M_{\rm Tot.}\,[\Msol])^{5}$ &   11.4  &  9.8 & $ 5.0\%$ \\  
				$\sfr [\Msolyr]^{6}$                   &   38.3  & 1.95 & $ 5.1\%$ \\
				$r [\kpc]^{7}$                         &    6    &  2   & $33.3\%$ \\ 
				\hline
				\hline
			\end{tabular}\\
			{\footnotesize
			Notes: {\bf Row 1:} gas mass, {\bf Row 2:} stellar mass,  {\bf Row 3:} total baryonic mass, {\bf Row 4:} dark matter mass, {\bf Row 5:} total mass, {\bf Row 6:} \Sfr for the last $10\Myr$ of runtime, {\bf Row 7:} Radius at time of analysis. 
			}
			\label{table:sprop}
		}	

\end{table}

\section{Galaxy Selection and Properties}
\label{sect:galaxydiscrip}

The majority of our analysis is carried out when simulation has run for $\sim3.4\Gyr$ (i.e. when the simulation has reached $z\sim 2$) by extracting a cubic volume, centred on our galaxy of choice. The sides of this volume all measure $30\kpc$ across. At $z=2$, the simulation has maximum spatial resolution of $\Delta x \sim45\pc$, however we carry out our analysis (unless otherwise stated) at one level below the maximum refinement (i.e. $\Delta x \sim90\pc$) to reduce the computation cost of when running \lcarsp. 
At runtime $\sim89\%$ (by mass) of the galaxy's dense gas ($n_{\rm H}\geq1\cc$)  is resolved to at least this resolution, with $\sim71\%$ being resolved to the maximum resolution. 
We rotate our selected galaxy before analysis so that its inclination angle ($i$) is $20^{\circ}$. We define the inclination angle so that $i=0^{\circ}$ corresponds to the angular momentum vector of the galaxy is pointing along the z-axis of the simulation and towards the reader. For $i\ne0$ the rotation occurs about the x-axis of the simulation, which is the same as the x-axis used in Fig.~\ref{fig:smaps}. 
\begin{figure}
	\begin{center}
		\includegraphics[width=0.4\textwidth]{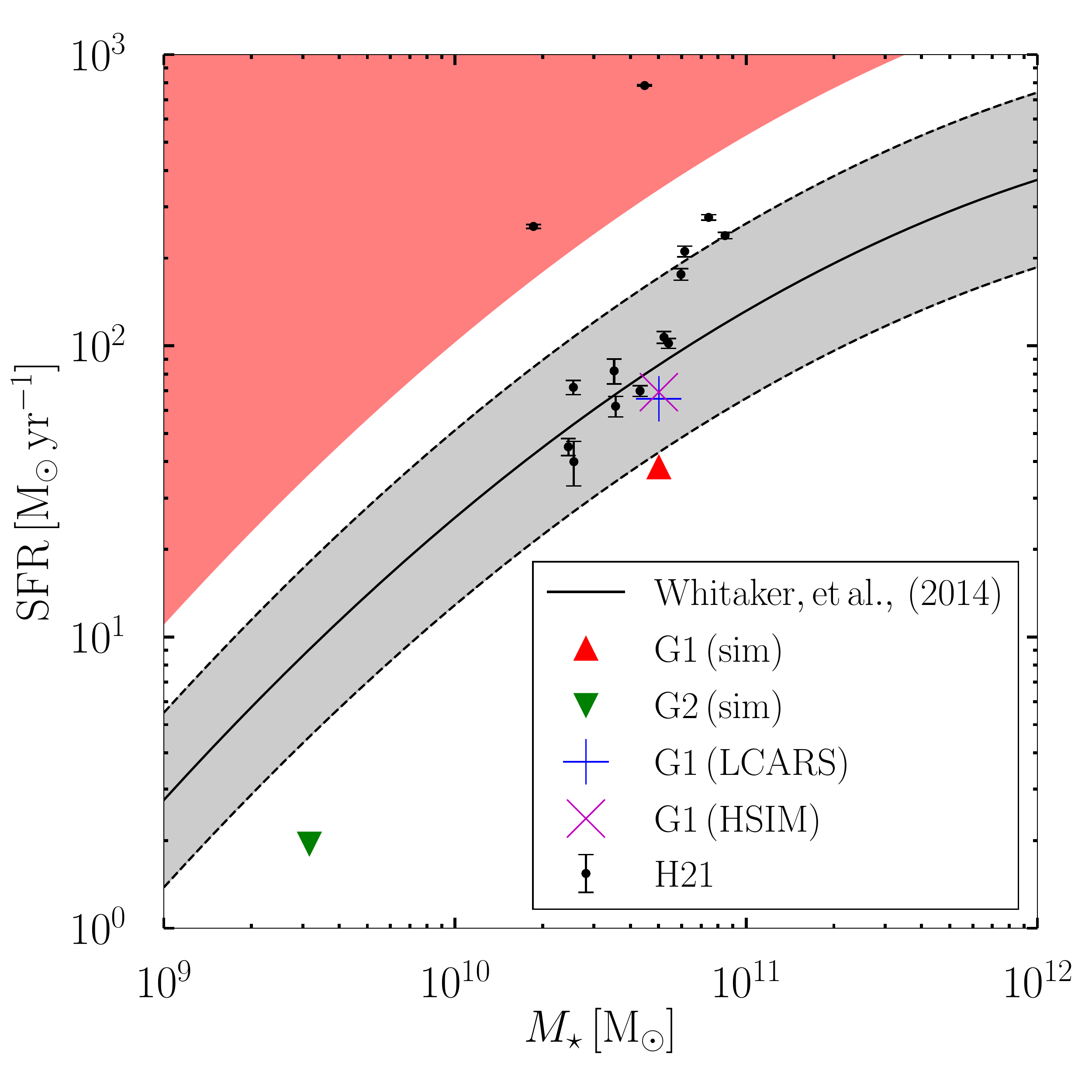} 
		\caption{The SFR position of G1 (red triangle) and G2 (green triangle) relative to the SFR-stellar mass main sequence as measured directly from the simulation. The black-solid line shows the main sequence for $2.0\leq z\leq2.5$  \citep[as described by Eq. 2 of][]{Whitaker:2014aa}. The grey shaded regions show values within $\pm0.3$ dex of the main sequence. The red shaded region shows where galaxies considered to be ``starburst'' are found. Observational \sfr\, calculated from IR luminosity data given in Table 3 of H21 are shown by black-circular points. The blue ``+'' and magenta ``x'' show the measured \sfr\, of G1 after being passed through \lcars and then \hsim (see \S\ref{res:hsim}) respectively.  
				}
	\label{fig:sms}
	\end{center}
\end{figure}
\begin{figure}
	\begin{center}
		\includegraphics[width=0.5\textwidth]{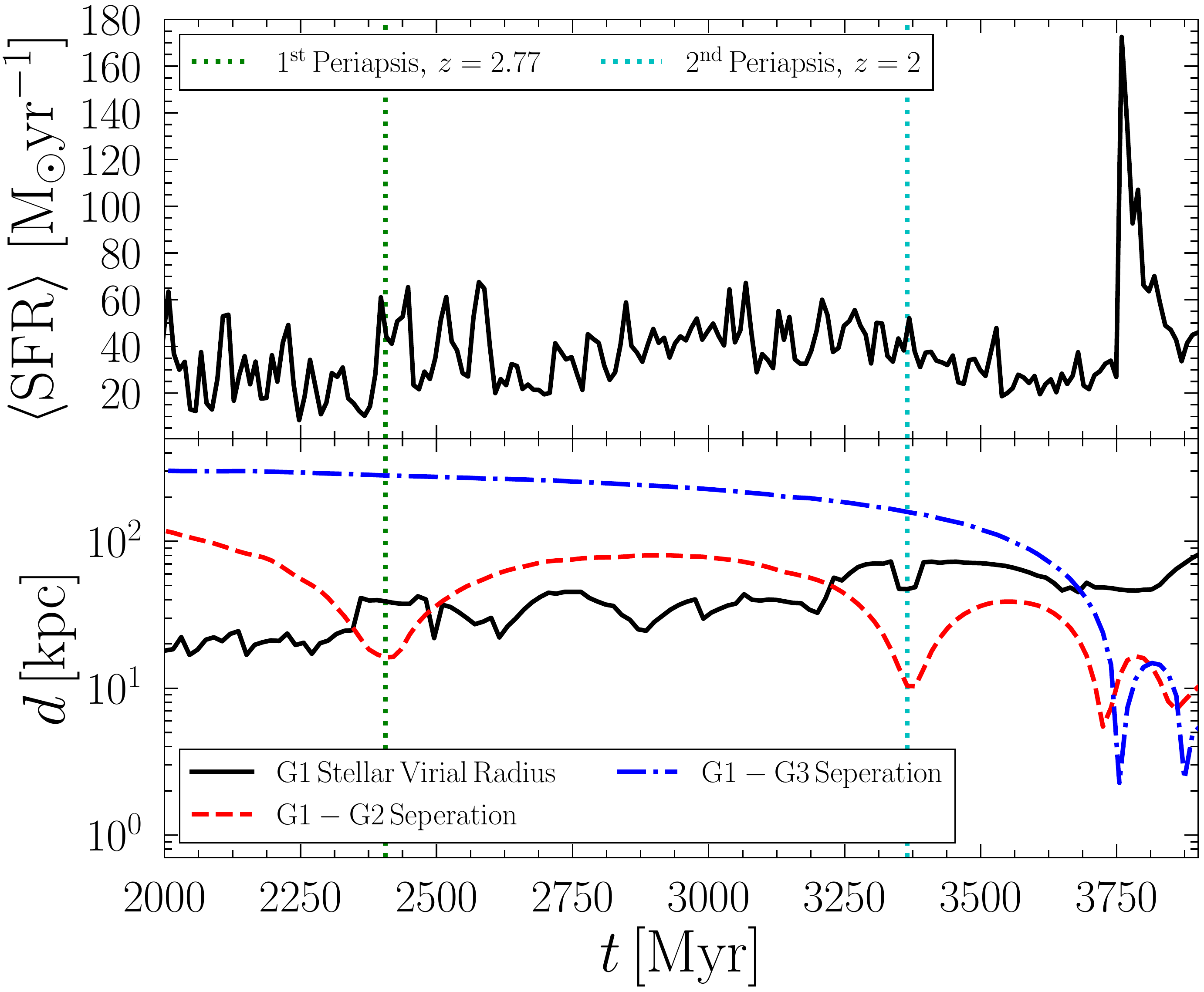} 
		\caption{Top: Evolution of the mean SFR per $10\Myr$ for G1 $\left(\langle\rm SFR\rangle\right)$. Bottom: Evolution of the seperation between G1 and its two satellite galaxies and the virial radius of G1's stellar component. The two vertical lines running through both panels indicate the first and second periapsis, with the latter being the time at which we analyse the simulation. }
	\label{fig:sfh}
	\end{center}
\end{figure}
\begin{figure*}
	\begin{center}
		\includegraphics[width=0.65\textwidth]{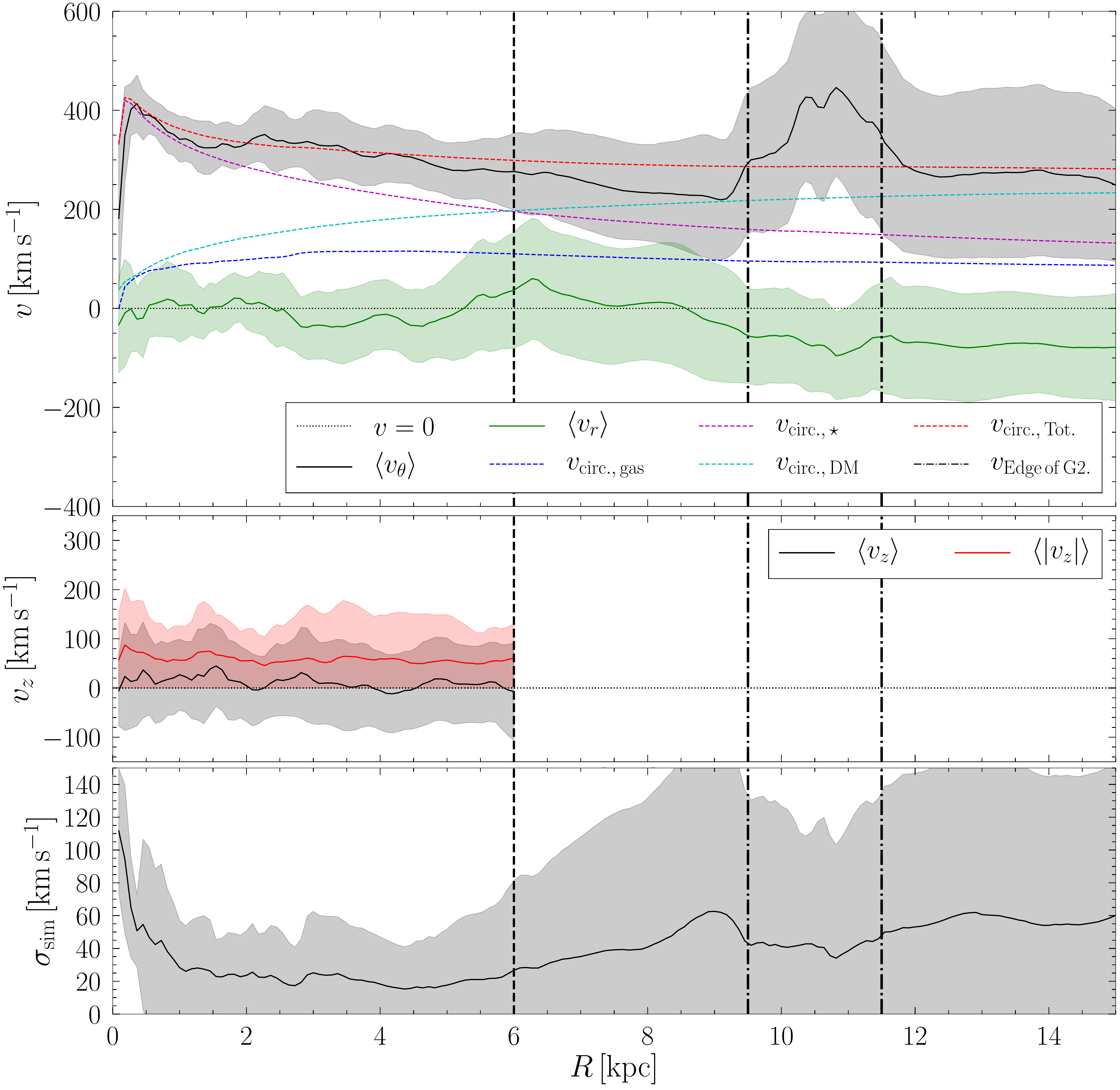} 
		\caption{
		Radial mean velocity profiles as a function of $R$. 
		Top: dashed lines show the expected circular velocity for each mass component: DM (cyan), stars (purple), gas (blue) and combined (red). The solid green and black lines show mean radial $\left(\langle v_{r}\rangle\right)$ and rotational $\left(\langle v_{\theta}\rangle\right) $velocities for the gas. 
		Middle: Gas velocity perpendicular to G1's disc ($v_{z}$). The black line shows the mass weighted average of $v_{z}$, while the red line shows the mass weighted average of the magnitude of $v_{z}$. By definition $v_{z}$ does not extend beyond edge of G1, therefore we do not show $v_{z}$ for $R>6\kpc$.
		Bottom: The mean radial profile of the velocity dispersion, $\sigmasim$, see \S\ref{lcars:spec} for details. 
		All: The shaded regions show $\pm$ one standard deviation from the mean.  The vertical dashed line represents the edge of G1, while the dot-dashed line shows the edge of G2.}
	\label{fig:vrot_actual}
	\end{center}
\end{figure*}

From \newhorizon we selected a galaxy who's properties are consistent with those of observed on the MS at $z\sim2$, i.e. has a stellar mass ($M_{\star}$) and \sfr\, consistent with galaxies at this redshift. We use the observational sample in our companion work H21 to determine a suitable range of values in $M_{\star}$ and \sfr.
Our selected galaxy is presented in Fig.~\ref{fig:smaps} and a summary of its intrinsic properties is given in Table~\ref{table:sprop}. From Fig.~\ref{fig:smaps} it is clear that our selected galaxy is interacting with a smaller galaxy. We refer to these two galaxies as G1 and G2 respectively. 
We note that it was not intentional to select a galaxy under going a merger. Given the ratio of their mass', $\sim5\%$ (see the third column of Table~\ref{table:sprop}) we classify the merger of G1 and G2 as minor.  
From a visual inspection, G1 appears to be spiral galaxy with two large arms and several small arms. In contrast G2 is comprised of a dense core with two circular gas filaments which extend above and below the galaxy.  We define young stars as those with ages less than or equal to $t_{\rm\star,y}=10\Myr$. We plot the positions of all young stars in Panel \fpiii\, of Fig.~\ref{fig:smaps} which shows that the majority of G1's star formation is occurring along the lengths of its arms, while G2 is predominantly forming stars in two central clusters. Comparing panels \fpii\, and \fpiii\, shows that older stars are more evenly distributed throughout both galaxies. 

G1 has a ${\rm SFR}$ of $\sim38.3\Msolyr$ at $z=2$, putting it within $\sim0.35{\rm\, dex}$ of the galaxy star formation MS as defined in \cite{Whitaker:2014aa}, see Fig.~\ref{fig:sms}. For direct comparison with observed $z=2$ galaxies we include the values calculated for 14 galaxies by H21. The evolution of G1's average ${\rm SFR}$ per $10\Myr$ $\left(\langle\rm SFR\rangle\right)$\footnote{The calculation of $\langle\rm SFR\rangle$ does not account for stars that were formed in G1 but subsequently ejected or remove stars that formed in a satellite of G1 before a merger. Therefore $\langle\rm SFR\rangle$ provides an estimate and the general trend of star formation at a given time rather than the precise value.} is shown in the top panel of Fig.~\ref{fig:sfh}. 
G1 has a very bursty star formation history. Such a bursty star formation rate is expected for the star formation prescription employed in \newhorizon \citep[see][]{Grisdale:2021aa}. At the time of our analysis $(z=2)$ G1 is undergoing a small burst in $\langle\rm SFR\rangle$, we note that despite this burst the galaxy is not a ``starburst'' galaxy. From the \newhorizon merge tree we are able to determine that at $z=2$, G2 has completed a single complete orbit of G1 and is currently at periapsis of its second orbit (see bottom panel of Fig.~\ref{fig:sfh}). At G2's first periapsis there is a noticeably larger increase in  $\langle\rm SFR\rangle$ than at second periapsis, which might suggest a larger impact cross section during the initial approach. At $t=3750\Myr$ the $\langle\rm SFR\rangle$ jumps to $\sim180\Msolyr$, from the merger tree and based on a visual analysis of the simulation we see that this corresponds to a near head on collision with a third galaxy (which we refer to as G3). After the initial interaction between G1 and G3, G3 is trapped both within the stellar viral radius and the stellar disc of the G1. In contrast it is not until after the second periapsis' that G2 is trapped within the stellar virial radius. In future work we will explore the impact of G3 on G1's evolution further. 
 
By determining the mass enclosed ($M(R)$) within a given galactic radius ($R$) for each of the constituent mass components (i.e. gas, stars, DM and BH) of the galaxy, their circular velocity $\left(v_{{\rm circ.},i}=\sqrt{GM_{i}(<R)/R}\right)$ is calculated and presented in Fig.~\ref{fig:vrot_actual}. Also calculated is $v_{{\rm circ.Tot.}}$, where $M_{\rm Tot.}(<R) = M_{\rm star}(<R) + M_{\rm gas}(<R) + M_{\rm DM}(<R) + M_{\rm BH}(<R)$. Comparing $v_{\rm circ.}$ of the various components reveals that for $R<6\kpc$ $v_{{\rm circ.Tot.}}$ is dominated by the stellar mass component of G1 while at larger $R$ DM is dominant. This change in dominant component is in excellent agreement with the visually assessed edge of G1's disc at $R=6\kpc$. The gas component of G1 has little impact on $v_{{\rm circ.Tot.}}$. Due to the near negligible effect that BH has on $v_{{\rm circ.Tot.}}$, $v_{\rm circ.,BH}$ is not included in the figure.

\begin{figure*}
	\begin{center}
		\includegraphics[width=1.0\textwidth]{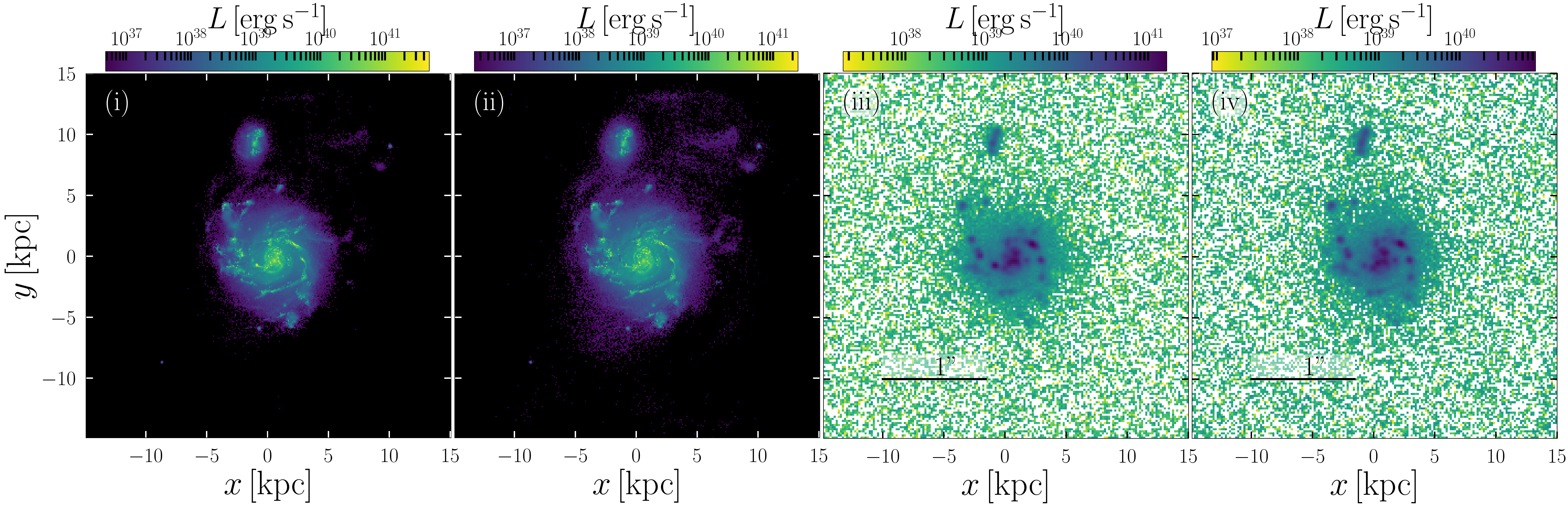} 
		\caption{${\rm H}{\alpha}$ $0^{\rm th}$ moment maps. The panels show: \fpi\, post-\lcarsp, pre-\hsimp, no extinction, \fpii\, post-\lcarsp, pre-\hsimp, with extinction. \fpiii\, post-\lcarsp, post-\hsimp, no extinction, \fpiv\, post-\lcarsp, post-\hsimp, with extinction. Continuum emission has not been subtracted so as to provide a clear indication of how such a galaxy is likely to appear before analysis. 
		}
	\label{fig:lcars_ha}
	\end{center}
\end{figure*}
Fig.~\ref{fig:vrot_actual} also includes the mean rotational ($\langle v_{\theta}\rangle$) and radial ($\langle v_{r}\rangle$) velocities\footnote{Both $v_{\theta}$ and $v_{r}$ are polar velocities and should not be compared with spherical velocities.} of the gas in G1 as a function of $R$. Within G1 (i.e. $R<6\kpc$) $\langle v_{r}\rangle$ is consistent with $0\kmsec$ (with gas moving both towards and away from the galactic centre) while $\langle v_{\theta}\rangle$ largely agrees with $v_{{\rm circ.Tot.}}$. This indicates that G1 is a rotationally supported disc as one might expect from a visual inspection of its morphology (see Fig.~\ref{fig:smaps}). In \S\ref{res:vsig} we attempt to quantitatively determine to what extent G1 is a  disc.
Outside of the G1, $\langle v_{\theta}\rangle$ shows a substantial ``bump'' between $R\sim9$ and $R\sim12\kpc$, this feature corresponds to G2. For $R\gtrsim9\kpc$ we find that $\langle v_{r}\rangle$ begins to decrease before stabilising at $\sim-69\kmsec$ which indicates that G1 is accreting gas from these larger radii (including G2). 
 
Completing the triplet of the cylindrical velocity components, the middle panel of Fig.~\ref{fig:vrot_actual} shows the (mass weighted) mean vertical velocity of G1. When accounting for the direction of gas movement, $\langle v_{z}\rangle$ tends to be between $0$ and $20\kmsec$. In a handful of radii gas moving out of the ``bottom'' of G1's disc is the more dominant direction of gas flow. Ignoring the direction of $v_{z}$ and just focusing on the magnitude, we find that $\langle |v_{z}|\rangle$ tends to $\sim60\kmsec$ at all radii. Therefore at any given radii, $v_{\theta}$ is the dominant velocity component of the vast majority of G1's gas. A visual inspection of G1 reveals the presence of (small) outflows and fountains which is not surprising, given the dispersion on both $\langle v_{z}\rangle$ and $\langle |v_{z}|\rangle$.
 
The mean radial profile of $\sigmasim$ is presented in the lower panel of Fig.~\ref{fig:vrot_actual}. For G1 we find the (mass weighted) mean $\sigmasim$ of $\sim23\kmsec$. This value excludes the central kiloparsec, which can have up to $\times5.4$ higher $\sigmasim$. The high values in the galactic centre are easily explained by the larger number of young stars and the galaxy's central massive black hole injecting energy in to the surrounding gas. At radii outside of G1 the mean of $\sigmasim$ increases to $\sim64\kmsec$, with some features in the profile around G2. 

Taking all of the above into account and combining it with visual inspections of the disc, we are able to conclude that the gas of G1 is gravitationally bound, but supported against collapse by galactic rotation. For individual regions within the galaxy, $v_{r}$, $v_{z}$ or turbulence ($\sigmasim$) can dominate over $v_{\theta}$ which can result in outflows, fountains and star formation. 


\begin{figure*}
	\begin{center}
		\includegraphics[width=1.0\textwidth]{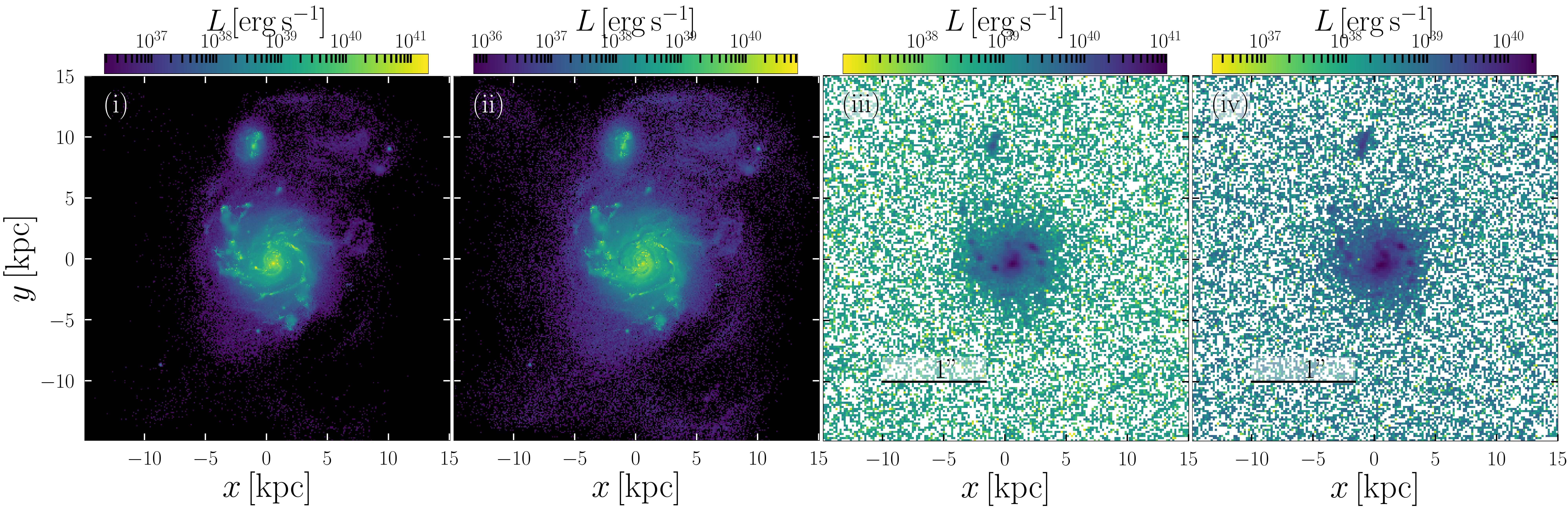} 
		\caption{${\rm H}{\beta}$ $0^{\rm th}$ moment maps. The panels show: \fpi\, post-\lcarsp, pre-\hsimp, no extinction, \fpii\, post-\lcarsp, pre-\hsimp, with extinction. \fpiii\, post-\lcarsp, post-\hsimp, no extinction, \fpiv\, post-\lcarsp, post-\hsimp, with extinction. Continuum emissions has not been subtracted so as to provide a clear indication of how such a galaxy is likely to appear before analysis. 
		}
	\label{fig:lcars_hb}
	\end{center}
\end{figure*}
\section{Results}
\label{sect:resultsi}
The following sections compare the galaxies intrinsic properties to those measured from the SSC created by \lcarsp, including the star formation rate and rotational velocity of the galaxy. 
We define ``intrinsic properties'' to be the values of a given property (e.g. stellar mass) of the galaxy measured directly from the simulation without adding photons, while ``observed properties'' are the value of the same property determined from mock observations after photons have been added by \lcarsp. For the purposes of this work the intrinsic properties are analogues to the physical properties that a galaxy in the physical Universe would have, while observed properties are those that would be derived from observations. 

At $z=2$ the wavelength of \ha falls edge of the K-band grating used by HARMONI. To avoid a loss of data we artificially move both G1 and G2 to $z=2.2$. This only changes the observed wavelength of the \ha emission line to $2.1{\,\rm\mu m}$ and the value of $D_{\rm L}$. This does not affect the results or conclusions drawn in this work.

\subsection{Post-\lcarsp, Pre-observation}
\label{res:lcars}

We begin our analysis by presenting the results of passing a $30\times30\times30\kpc^{3}$ volume from \newhorizon, centred on G1, through \lcarsp. As stated in \S\ref{sect:galaxydiscrip}, the extracted volume is passed to \lcars with a fixed spatial resolution of $\Delta x\sim90\pc$ and results in \lcars running $624,729$ unique \cloudy simulations. Of these $408,259$ are for cells containing SSPs while the remaining $216,470$ are for cells contaminated by diffuse emission (see \S\ref{lcars:diffuse}). 

Throughout this section we focus on the ability to recover the G1's intrinsic properties from the output of \lcars. We leave a discussion of the impact of the telescope (i.e. \eelt and HARMONI) on the measured properties until \S\ref{res:hsim}. 

\subsubsection{\ha Morphology}
\label{res:lcars:morph}
From the \lcars SSC it is possible to calculate the $0^{\rm th}$ moment, i.e. $L=\int I(\lambda)d\lambda$, of the \ha emission at each spatial pixel location and thus produce the maps shown in panels \fpi\, and \fpii\, of Fig.~\ref{fig:lcars_ha} which are centred on the \ha emission line. The first of these maps shows the galaxy without any extinction being applied. The post-\lcars maps appear to be an amalgamation of G1's gas and stellar structures. In both maps the spiral arms are clearly visible and brighter than the more diffuse gas of the disc. Comparing either map from Fig.~\ref{fig:lcars_ha} with the distribution of young stars (i.e. those with ages $\leq t_{\rm\star,y}$) in the simulation (see Fig.~\ref{fig:smaps}, panel \fpiii) shows that the \ha emission is an excellent tracer of young stars, as expected. 

As one might expect, extinction has significant impact on the brightness of G1's spiral arms. As the spiral arms are the primary sites of star formation (other than the galactic centre, see Fig.~\ref{fig:smaps}) these are the regions where the brightest pixels are found in the SSCs. The net result of reducing the brightness of the arms is that the diffuse gas between G1 and G2, as well as the gas that is extending towards the right of the maps, \emph{appears} brighter compared to the arms.  Fig.~\ref{fig:lcars_hb} shows integrated emission maps for the \hb emission line. Just as with \ha the strongest emission in \hb comes from the spiral arms. We note that  \hb map experiences more extinction than \ha map, as expected.

\subsubsection{Measured Star Formation Rates}
\label{res:lcars:sfr}

Using a conversion factor it is possible to calculate the star formation rate of a galaxy from its \ha line emission  \citep[][]{Kennicutt:1994aa}. In this work, as in H21, we adopt the conversion
\begin{equation}
	{\rm SFR_{\rm H\alpha}}\, [{\Msolyr}] = 5.37\times10^{-42}L_{\rm H\alpha}\,[\ergsec],
	\label{eq:sfrha}
\end{equation}
from \cite{Murphy:2011aa}. Here $L_{\rm H\alpha}$ is the integral of the continuum subtracted, single aperture, emission line for the entire galaxy. Using Eq.~\ref{eq:sfrha} with the \ha emissions measured from the SSC without extinction we calculate ${\rm SFR_{\rm H\alpha}}=59.1\,\Msolyr$. This is $\sim1.54\times$ the actual SFR of the galaxy.

To calculate the SFR from the \ha SSC with extinction included we first need to correct for the extinction using 
\begin{equation}
	L_{\rm H\alpha, emitted} = L_{\rm H\alpha, observed}\cdot10^{0.4A_{\rm H\alpha}},
	\label{eq:ext_corr1}
\end{equation}
where $A_{\rm H\alpha}=(3.33\pm0.8)E(B-V)$ and  
\begin{equation}
	E(B-V) = \delta k\log_{10}\left( \frac{(L_{\rm H\alpha}/L_{\rm H\beta})_{\rm observed}}{2.86}\right),
	\label{eq:ext_corr2}
\end{equation}
see \cite{Dominguez:2013aa}, \cite{Osterbrock:1989aa} and references therein. The pre-factor $\delta k$ is set by the choice of extinction curve and for consistency we use the same Fitzpatrick curve employed by \lcars (see \S\ref{lcars:spec}) which gives $\delta k=1.50$.  Integrating a single aperture spectrum for both \ha and \hb from their respective  SSC (which include extinction), $L_{\rm H\alpha}$ and $L_{\rm H\beta}$ can be calculated and passed through equations Eq.~\ref{eq:ext_corr1} and \ref{eq:ext_corr2} which gives $A_{\rm H\alpha}\sim0.75$. We then calculate ${\rm SFR_{\rm H\alpha}}=65.8\,\Msolyr$ using Eq.~\ref{eq:sfrha}, which is $\sim1.72\times$ larger than the intrinsic \sfr\, of the galaxy. The factor of $\sim1.5-1.7$ difference between the intrinsic and measured \sfr\, could come from the choice of \ha to \sfr\,conversion factor, choice and application of the extinction curve, etc. Given that the measured \sfr\, values do not move G1 off of the MS (see Fig.~\ref{fig:sms}), we argue that these differences are acceptable and that we are able to recover, at least to an order of magnitude the correct \sfr\, of G1.  
We note that in above calculations only cells that are within a radius of $6\kpc$ from the galactic centre are considered. 

\subsubsection{Kinematics Structures}
\label{res:lcars:kinem}
\begin{figure}
	\begin{center}
		\includegraphics[width=0.5\textwidth]{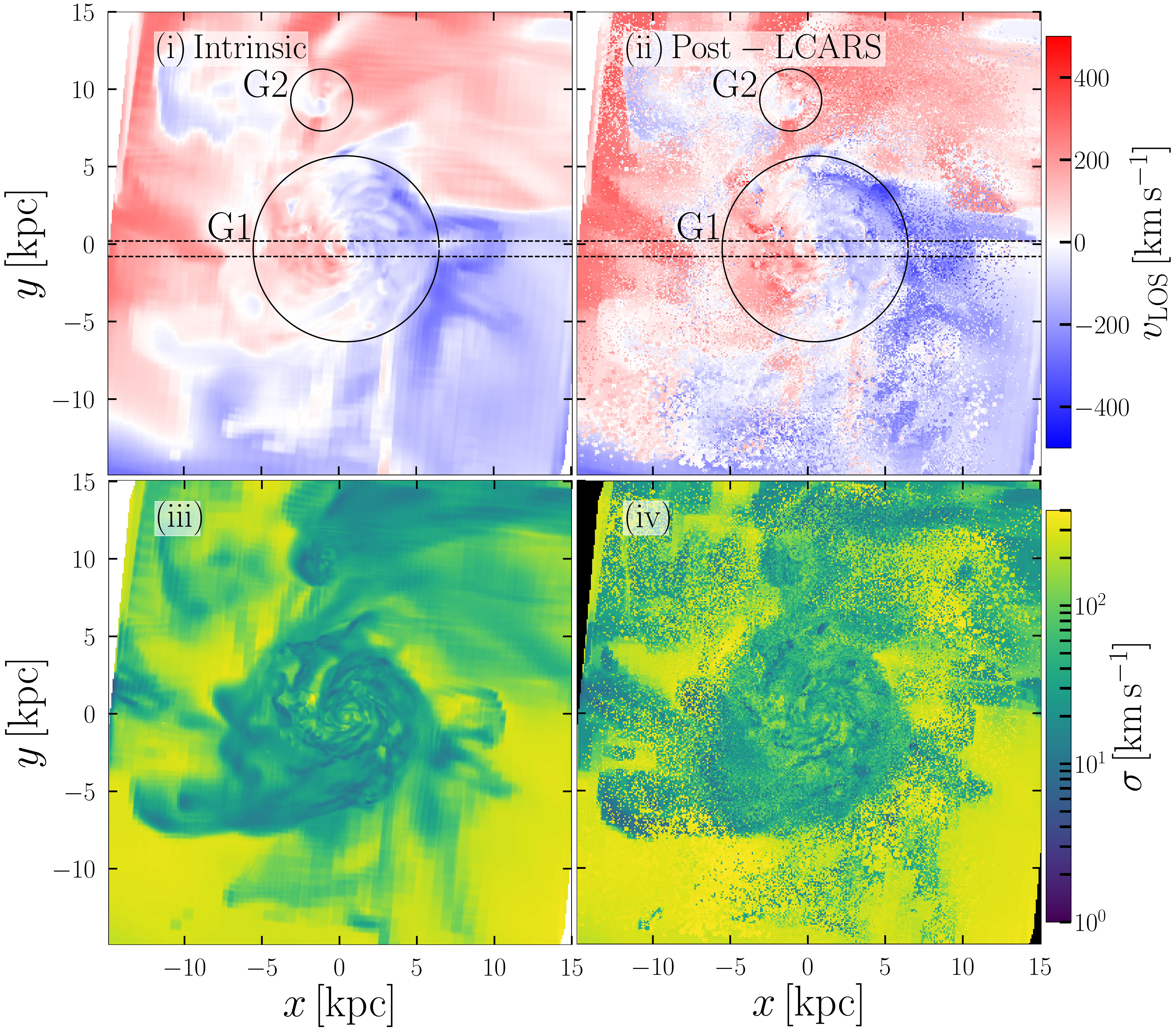} 
		\caption{Comparison of intrinsic and observational velocity Maps of G1 and G2. Panels \fpi\, and \fpiii\, show the intrinsic line of sight velocity and velocity dispersion ($\vlosma$ and $\sigmma$ respectively). Panels \fpii\, and \fpiv\, show the $1^{\rm st}$ and $2^{\rm nd}$ moment maps ($\vlosmi$ and $\sigmaobs$ respectively) for the SSC produced by \lcars. Continuum is subtracted before $\vlosmi$ and $\sigmaobs$ are calculated. To aid in comparisons, panels on the same row use the same colour scale which is shown on the right of row. See \S\ref{res:lcars:kinem} for definitions of quantities. The dashed horizontal lines show the edges of the slit used for $v_{\theta,{\rm slit}}$ calculations. 
		}
	\label{fig:kinematics_1}
	\end{center}
\end{figure}
\begin{figure}
	\begin{center}
		\includegraphics[width=0.5\textwidth]{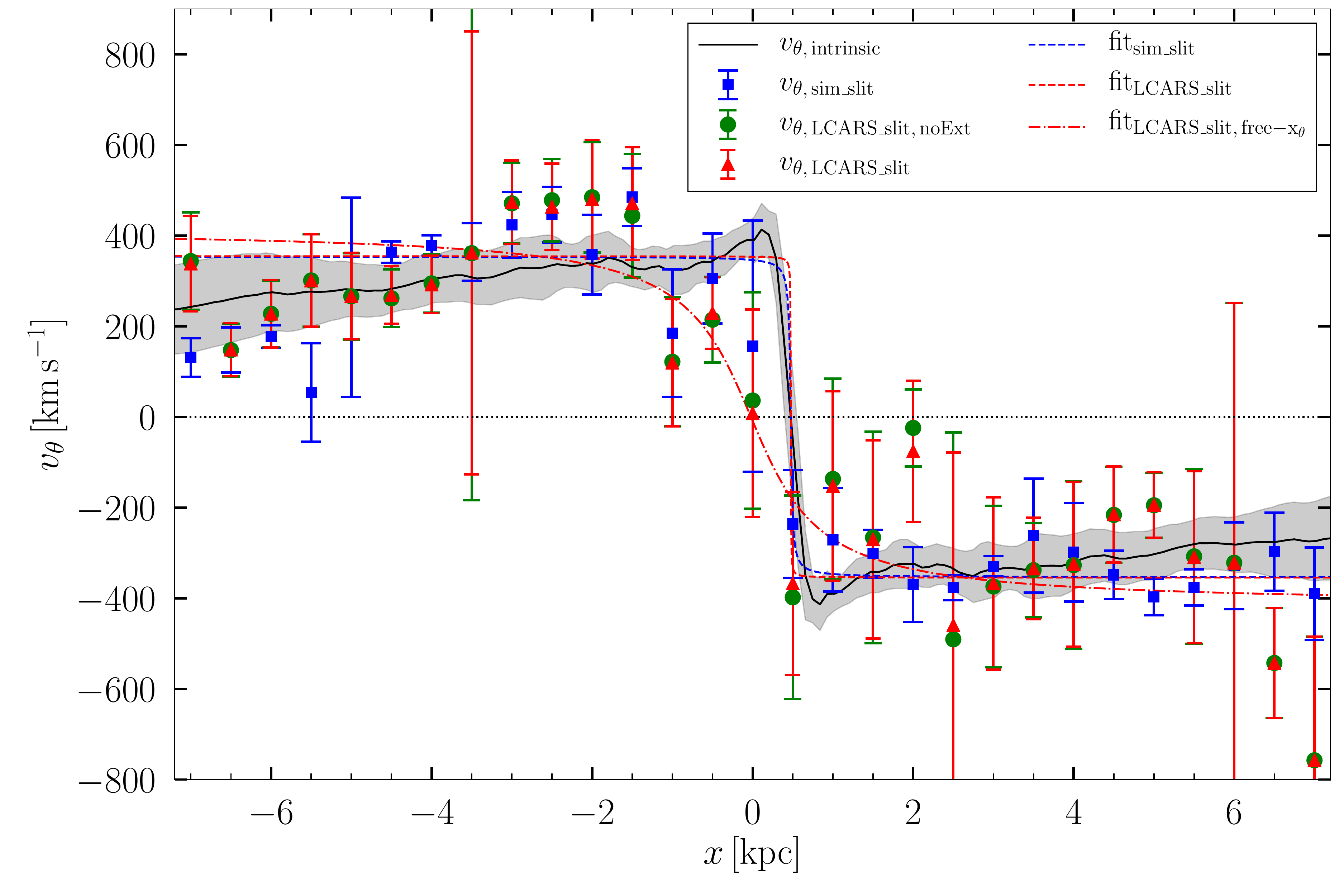} 
		\caption{$v_{\theta}$ profiles calculated from pixels extracted from the slit shown in Fig.~\ref{fig:kinematics_1}. The intrinsic rotation curve (solid black line) is included to aid in comparison. The coloured data points represent the weighted mean value of $v_{\theta}$ calculated from the slit every $0.5\kpc$, while their error bars show the associated weighted $1\sigma$ value. Blue, magenta and red data points show values calculated from: the $\vlosma$ map, the $\vlosmi$ map neglecting extinction and the $\vlosmi$ map respectively. The dashed lines represent an arctan fit to data points with matching colour. The dash-dot line shows an arctan fit to the $\vlosmi$ map, where position of $x_{\theta}$ is not known. 
		Note that $x=0\kpc$ is centre of the map \emph{not} the kinematic centre of the galaxy.
		}
	\label{fig:kinematics_2}
	\end{center}
\end{figure}
In panel \fpi\, of Fig.~\ref{fig:kinematics_1} we show the \emph{mass} weighted mean value of $\vlos$ along the line of sight ($\vlosma$) map of G1 and G2. Some of the G1's spiral structure can be seen in the $\vlosma$ maps, though less clearly. From this map we find that in general, $|\vlosma|$ is almost always $<500\kmsec$ and that the lower density gas found at $R>6\kpc$ appears to be the most energetic. We also see that the top of G2 is moving way from the observer while the bottom is moving towards them: i.e. G2 appears to be rotating in the $y-z$ plane. 

In order to determine $\vlos$ post-\lcars we employ the observational method of taking the $1^{\rm st}$ moment of the continuum subtract spectrum in each cell, i.e.
\begin{equation}
	\vlosmi =\frac{\int v_{\lambda} I(v_{\lambda}) \diff v_{\lambda}} {\int I(v_{\lambda})\diff v_{\lambda}},
	\label{eq:mom1}
\end{equation}
where $v_{\lambda}$ is related to wavelength via Eq.~\ref{eq:lshift} and $I(v_{\lambda})$ is the intensity in the corresponding velocity (wavelength) channel. The $1^{\rm st}$ moment map is shown in panel \fpii\, of  Fig.~\ref{fig:kinematics_1}. Within G1 ($R\lesssim6\kpc$) the $1^{\rm st}$ moment map agrees with $\vlos$ map calculated directly from the simulation, in particular large scale structures match e.g. the gas on the bottom-right side of the galaxy in both maps is moving towards the observer. On smaller scales, within G1, the velocity structures are similar but with some differences in the fine structure (i.e. on scales of $<0.5\kpc$). Perhaps one of the most important difference being that $|\vlosmi|\sim500\kmsec$ is found within the disc of G1 but not in $\vlosma$. We discuss the source of these high velocity regions in the $1^{\rm st}$ moment map in \S\ref{dis:weighting}. Due to a lack of young, bright, \Ha-emitting stars, the spectrum for a given pixel at $R\gtrsim6\kpc$ is continuum dominated which results in the noisy measurements of $\vlosmi$ shown in Fig.~\ref{fig:kinematics_1}

The spiral structure of the G1 is more easily seen in panel \fpiii\, of Fig.~\ref{fig:kinematics_1}, which shows the \emph{mass} weighted mean value of $\sigmasim$ (as defined in Eq.~\ref{eq:gasbroad0}) along the line of sight ($\sigmma$) map. This map recovers a lot of the detailed gas structure seen in the gas surface density maps of the galaxy (see Fig.~\ref{fig:smaps}). The $2^{\rm nd}$ moment,
\begin{equation}
	\sigmaobs = \sqrt{\frac{\int I(v_{\lambda})(v_{\lambda}-\vlosmi)^2\diff v} {\int I\diff v_{\lambda}}},
	\label{eq:mom2}
\end{equation}
provides the observational equivalent of $\sigmma$, though as with $\vlosmi$ this is photon weighted. In panel \fpiv\, of Fig.~\ref{fig:kinematics_1} we show a map of $\sigmaobs$. Despite being noisier than the $\sigmma$ map the two maps show broadly the same structure. For example, both maps show that the spiral arms are regions of fairly uniform motion.  Some of the details in $\sigmaobs$ is lost due to noise, i.e. spaxel's with strong continuum relative to the emission. 

As the galaxy is inclined at $i=20^{\circ}$, $\vlos$ contains contributions from both $v_{\theta}$ and $v_{r}$ as well as $v_{z}$. As the position of the emission line for each cell is set by $\vlos$ (see \S\ref{lcars:spec}) $\vlosma$ also contains contributions from the $v_{\theta}$ of each of the cells combined along the line of sight. We apply a slit across the $\vlosmi$ map (shown as dashed lines in Fig.~\ref{fig:kinematics_1}). The slit is positioned to run through the centre of G1 and along the inclination axis. For each pixels within this slit we approximate $v_{\theta}$ from $\vlosmi$ as
\begin{equation}
	v_{\theta,{\rm slit}} = \frac{\vlosmi}{\sin(i)}.
	\label{eq:vlos2vrot}
\end{equation}
By taking the weighted mean and standard deviation ($1\sigma$) values of $v_{\theta,{\rm slit}}$ every $0.5\kpc$ along the $x$-axis we are able to produce estimates of the rotation curve. We apply the same slit method to the $\vlosma$ map to provide a comparison. In the former case pixels are weighted by their integrated line emission, while the latter is weighted by the gas mass of a pixel. The resulting rotation curves are presented in Fig.~\ref{fig:kinematics_2}.

We check the validity of calculating the rotations curves from a slit by comparing the slit-calculated curve from $\vlosma$ to the intrinsic $v_{\theta}$ curve,\footnote{To aid with the comparison we mirror and flip the intrinsic curve to produce an $\arctan$-like curve which covers the width of the entire slit. Additionally, the curve is shifted horizontally so that $v_{\theta,\,{\rm intrinsic}}=0.0$ at the kinematic centre of the galaxy which is found at $x=0.475\kpc$.} and we denote those two rotation curves as $v_{\theta,\,{\rm sim\_slit}}$ and $v_{\theta,\,{\rm intrinsic}}$ respectively. An exact match between the two is not expected given that $v_{\theta,\,{\rm intrinsic}}$ is calculated for a significantly larger data set, i.e. the entire galaxy, rather than just a narrow slice through the centre.\footnote{19,243,854 individual cells are used when calculating the $v_{\theta,\,{\rm intrinsic}}$ curve compared to the 2,497 pixels used in the slit-calculated method.} This means that a small, (abnormally) high/low-velocity region which crosses the slit will have a more substantial impact on $v_{\theta,\,{\rm sim\_slit}}$ than it would have if all cells within that galaxy at that particle radius were included in the calculation.
With that said, except for a handful of data points, when $|x|\leq6\kpc$ we find that $v_{\theta,\,{\rm sim\_slit}}$ is within $1\sigma$ of the intrinsic value and therefore consistent with $v_{\theta,\,{\rm intrinsic}}$. The data points that deviate can easily be explained by small scale structure seen in panel \fpi\, of Fig.~\ref{fig:kinematics_1}. 
We conclude that the slit calculated curves provide a reasonable estimate of the G1 rotational velocity.

We now turn our attention to the rotation curve calculated from the $\vlosmi$ map, $v_{\theta,\,{\rm LCARS\_slit}}$. It could be argued that all data point, except the one at $x=2\kpc$, are consistent with $v_{\theta,\,{\rm intrinsic}}$ (when $1\sigma$ are considered). However this is in part due to the large values of $1\sigma$ found for $v_{\theta,\,{\rm LCARS\_slit}}$. The $\vlosmi$ map in Fig.~\ref{fig:kinematics_1} shows a larger range of values within G1 than seen in the $\vlosma$ map. For example at $(-2.9,-0.25)$ we see a small (blue) region moving in the opposite direction to its surroundings, which provides an explanation for $1\sigma\sim488\kmsec$ on $v_{\theta,\,{\rm LCARS\_slit}}$ at $x=-3.5\kpc$. Furthermore, for $R>6\kpc$ the large amount of noise in $\vlosmi$ leads to large uncertainty on $v_{\theta,\,{\rm LCARS\_slit}}$ at these radii.

By fitting an arctan profile to both $v_{\theta,\,{\rm sim\_slit}}$ and $v_{\theta,\,{\rm LCARS\_slit}}$ we are able to quantitatively compare how processing G1 with \lcars has impacted the recovered rotation curve. 
When fitting we provide the x-position of the centre of rotation, $x_{\theta}=0.475\kpc$, which is calculated directly from the simulation.
 We find the asymptotes of the arctan fit to be at $354\pm5$ and $355\pm22\kmsec$ respectively. Both values are a reasonable match to $v_{\theta,\,{\rm intrinsic}}$ at its flattest part ($0.5\lesssim R\lesssim3.0\kpc$), which we find to be $\sim340\kmsec$. The fitted arctan profile also captures the sharp increase in $v_{\theta,\rm intrinsic}$ found around the galactic centre. 

Assuming that $v_{\theta,\,{\rm LCARS\_slit}}=v_{{\rm circ.Tot.}}$ (at $R=6\kpc$), we are able to estimate the total mass (i.e. sum of stellar, gas, dark matter \& black holes mass) enclosed within G1's disc as $1.76\cdot10^{11}\Msol$ from the \lcars \ha SSC. This is $\sim1.39\times$ larger than the actual total mass of G1, see Table.~\ref{table:sprop}. Thus to within a factor of a few we are able to recover the total mass of the G1. The steady decrease in the rotation curve seen at $R\gtrsim3\kpc$, which is not modelled by an arctan profile, provides the explanation for the difference between the intrinsic and measured enclosed mass at $R<6\kpc$.

If $x_{\theta}$ is instead left as a free parameter, the fitted arctan profile does not accurately capture the galactic centre and leads to a larger value of $v_{\theta}$ for the asymptote, i.e. $416.3\pm27.2\kmsec$.  Estimating the mass of the galaxy from this fit gives an enclosed mass of $2.42\cdot10^{11}\Msol$ at $R<6\kpc$ (i.e. $1.92\times$ the true value). Therefore, even if the galactic centre isn't known a reasonable, but higher, values of both $v_{\theta}$ and the enclosed mass can still be estimated. 

In summary, the analysis shows that creating SSCs using \lcars encodes rotational velocity information about galaxy into the emission lines as intended. 
With prior knowledge of the galaxy (such as inclination or galaxy centre) and an absence of observational effects (e.g. noise, sky lines, etc) the above slit method can be used to extract the rotational curve from a galaxy with reasonable accuracy. Therefore we argue that this method is useful for testing if $v_{\theta}$ has been encoded in the outputs of \lcars but caution its use for actual observational data. In \S\ref{hsim:kinematic} we explore an alternative method which can be applied to observational data.

\subsection{Post-\lcarsp, Post-observation}
\label{res:hsim}

\begin{figure*}
	\begin{center}
		\includegraphics[width=0.9\textwidth]{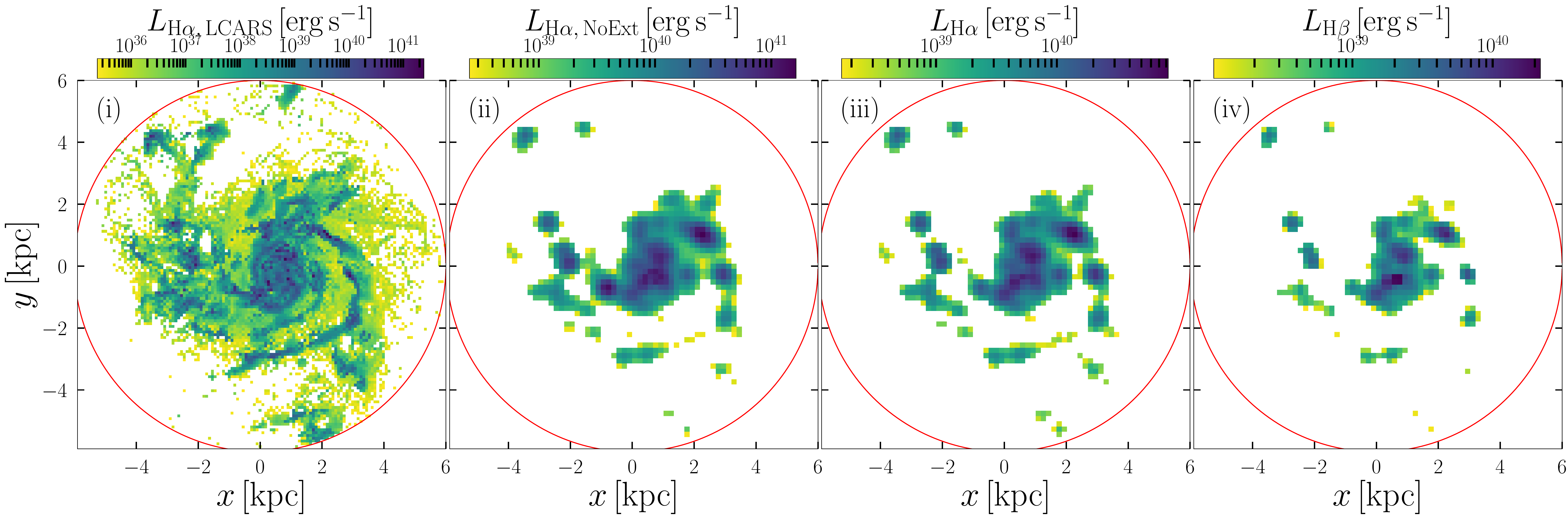} 
		\caption{$0^{\rm th}$ moment maps after the application of a SNR and radius mask have been applied. Panels \fpii-\fpiv\, show maps created from \hsim output data cubes of: \ha (without extinction), \ha (with extinction) and \hb (with extinction) respectively. For these panels only spaxels with ${\rm SNR}\geq3$ are included in the maps. Panel \fpi\, is the same as panel \fpii\, in Fig.~\ref{fig:lcars_ha} but zoomed in. The red ring in each panel represents edge of G1 and the radius of the radial mask. Continuum emission has been subtracted from all four panels.
		}
	\label{fig:hsim_msk}
	\end{center}
\end{figure*}
The analysis in the previous section was all carried out without accounting for a telescope or spectrograph. In the following section we pass the SSCs from \lcars through \hsim to produce mock ELT observations of G1. This simulator models all \emph{known} instrumental effects introduced by the HARMONI \citep[we direct the reader to][for an overview of \hsimp]{Zieleniewski:2015aa}. In this work we employ \hsim v3.03.\footnote{Details of this version can be found at \url{https://github.com/HARMONI-ELT/HSIM}} We summarise our default \hsim ``observation'' settings in Table~\ref{table:hsim}. These settings are adopted for all ``observations'' unless stated otherwise. As the primary goal of this work is to explore which physical values of a galaxy can be recovered from observations with HARMONI on the \eelt (and by extension other $30{\,\rm m}$-class telescopes) the observation settings used have been selected to provide the best possible data from an observation with ideal observational conditions (e.g. no moon). 
We leave exploring how different telescope settings will impact the accuracy of observations to future work.

In this work we use the ``\texttt{reduced}'' \hsim output cubes, which contain the simulated observations (with noise) after sky observations have been subtracted. Additionally we use the accompanying signal-to-noise ratio (SNR) cube ``\texttt{reduced\_SNR}'' as a means of filtering out spaxel's dominated by noise rather than signal from G1. 
\begin{table}

		\parbox{0.5\textwidth}{	
			\caption{{\sc hsim Settings}}
			\begin{tabular}[h]{l r }		
				\hline \hline
				Parameter                                 & Settings     \\
				                                          &              \\
				\hline
				Exposure Time (s)                         &    900       \\  
				Number of Exposures (without Extinction)  &    20        \\
				Number of Exposures (with Extinction)     &    40        \\
				Total Exposures time (without Extinction) &     5 hours  \\
				Total Exposures time (with Extinction)    &    10 hours  \\
				Spatial Pixel Scale (mas)                 &    $20$x$20$ \\
				Adaptive Optics Mode                      &    LTAO      \\
				Zenith seeing ('')                        &    0.57      \\
				Air Mass                                  &    1.1       \\
				Moon Illumination                         &    0.0       \\
				Telescope Jitter sigma (mas)              &    3.0       \\
				Telescope Temperature (K)                 &    280       \\
				Atmospheric Differential Refraction       &    True      \\
				Noise Seed                                &    10        \\
				Grating (\ha)                             &    K         \\
				Grating (\hb)                             &    H         \\
				\hline
				\hline
			\end{tabular}\\
			\label{table:hsim}
		}	

\end{table}

\subsubsection{Morphology and Star Formation Rates}
\label{hsim:morph}

 Panels \fpiii\, and \fpiv\, of Fig.~\ref{fig:lcars_ha} show the $0^{\rm th}$ moment map of the \ha emission line after G1 has been observed with \hsimp.\footnote{We have not employed a SNR filter to Fig.~\ref{fig:lcars_ha} so as to provide the reader with a clear expectation of the expected noise when observing a G1-like galaxy with the physical HARMONI on the \eelt.} 
 While the centre of the galaxy and the two largest spiral arms are still clearly visible, the majority of the structural detail seen in \fpii\, has now been lost. Even when extinction is neglected, i.e. panel \fpiii, these details are not recovered. Post-\hsim  G1 also appears smaller, with an apparent radius of $\sim4\kpc$. G2 is still discernible however it is greatly reduced in size. \hb observations (see Fig.~\ref{fig:lcars_hb}) show similar change but to a larger degree, i.e. G1's apparent size is further reduced and G2 is effectively lost in noise. We stress that compared to current observations the level of structure recovered by observations with \hsimp, and thus should be recovered with HARMONI, are still significantly better than what is achievable with current telescopes, see the discussion in \S\ref{dis:resolution2}.
 
We apply a SNR mask to the \hsim outputs to reduce the impact of noise on our measurements of ${\rm SFR_{\rm H\alpha}}$, i.e. all spaxels with a SNR$<3$ are masked. As in \S\ref{res:lcars:sfr}, only spaxels that are within $6\kpc$ of galactic centre (i.e. only spaxels belonging to G1) are included in ${\rm SFR_{\rm H\alpha}}$ calculations. Figure~\ref{fig:hsim_msk} shows the line emission maps after the two masks have been applied. Again taking a single aperture spectrum, carrying out continuum subtraction and applying Eq.~\ref{eq:sfrha} we recover a ${\rm SFR_{\rm H\alpha}}=50.9\pm0.0009\Msolyr$
from the extinction free \ha SSC. This value is below the value we measure from the \ha SSC before observation with \hsim ($\sim59.1\Msolyr$). 

Comparing panels \fpi\, and \fpii\, of Fig.~\ref{fig:hsim_msk} provides an explanation for this discrepancy: the more extended, inter-arm emissions of G1, is below the SNR cut. Little to no star formation is occurring in the inter-arm spaces (see panel \fpiii\, of Fig.~\ref{fig:smaps}), however due the diffusion module of \lcars (see \S\ref{lcars:diffuse}) these regions contain photons emitted by star forming regions and thus have integrated \ha luminosities between $\sim10^{38}$ and $\sim10^{39}\ergsec$ which is two or more magnitudes lower than the star forming arms.

We now turn to the SFR for the SSCs which include extinction. As in \S\ref{res:lcars:sfr} the \ha flux needs to be corrected and again we use the \hb emission line to estimate the correction. After applying the same SNR and radial masks as above, followed by Eq.~\ref{eq:ext_corr1} and \ref{eq:ext_corr2}, we compute a single aperture spectrum from which we calculate an $A_{\rm H\alpha}\sim1.036\pm0.0001$ and thus a ${\rm SFR_{\rm H\alpha}}=69.3\pm0.0003\Msolyr$. This measured \sfr\, is $\sim1.81\times$ the intrinsic \sfr. This value is within $6\%$ of the value obtained directly from the \lcars SSC. We are therefore able to conclude that \eelt and HARMONI systematics shouldn't significantly impact the measurement of \sfr\, and that the method used to convert \ha emissions to \sfr\, will be the most important source of any errors in observations.  

The uncertainties we report for the observed $A_{\rm H\alpha}$ and ${\rm SFR_{\rm H\alpha}}$ are calculated by propagating the variance cube (i.e. the inverse of the ``\texttt{reduced\_SNR}'' data cube) through Eq.~\ref{eq:sfrha}, \ref{eq:ext_corr1} and \ref{eq:ext_corr2}. As mentioned previously, we chose to employ the optimal observing conditions for kinematic analysis when carrying out our mock observations. Voxels above the SNR cut which contain signal from an emission line are typically several orders of magnitude stronger than any noise or variance. Consequently, we find very small uncertainties on for both $A_{\rm H\alpha}$ and ${\rm SFR_{\rm H\alpha}}$. In typical observations, the conditions are likely to be less optimal and therefore higher uncertainties should be expected.

 The choice of SNR cut can have an impact on the value of $A_{\rm H\alpha}$ which in turn impacts measured \sfr. For example, reducing the SNR limit to $2.0$ results in a larger $L_{\rm H\beta}$ while $L_{\rm H\alpha}$ remains approximately constant, this produces a smaller $A_{\rm H\alpha}(\sim0.72)$ and thus $\rm SFR_{\rm H\alpha}=55.89\pm0.0006\Msolyr$. However, the choice of SNR cut is not as important as how extinction is accounted for and how \ha is converted to \sfr.  
In summary, it is possible to recover the SFR of G1 after observations in within a factor of $\sim2$. 

\subsubsection{Kinematic Structures}
\label{hsim:kinematic}

\begin{figure}
	\begin{center}
		\includegraphics[width=0.5\textwidth]{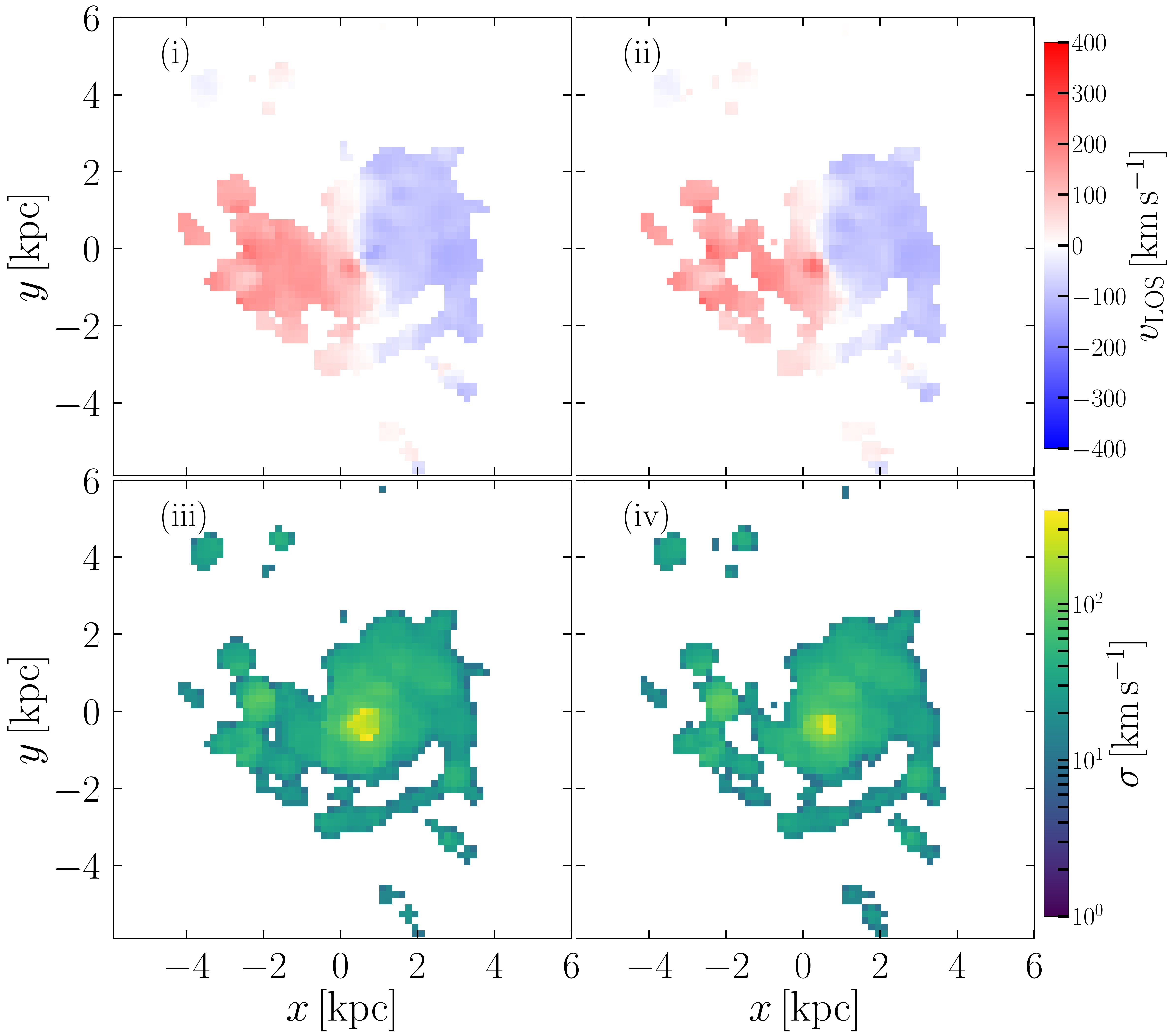} 
		\caption{$1^{\rm st}$ and $2^{\rm nd}$ moment maps ($\vlosmi$ and $\sigmaobs$ respectively) after our \ha SSC have been observed with \hsimp. Panels \fpi\,and \fpii\,show $\vlosmi$ for SSC without and with extinction, respectively, while panels \fpiii\,and \fpiv\, show there respective $\sigmaobs$.
		Continuum is subtracted and a SNR mask is applied before $\vlosmi$ and $\sigmaobs$ are calculated. To aid in comparisons, panels on the same row use the same colour scale which is shown on the right of row. See \S\ref{res:lcars:kinem} for definitions of quantities. 
		}
	\label{fig:kinematics_3}
	\end{center}
\end{figure}
\begin{figure}
	\begin{center}
		\includegraphics[width=0.5\textwidth]{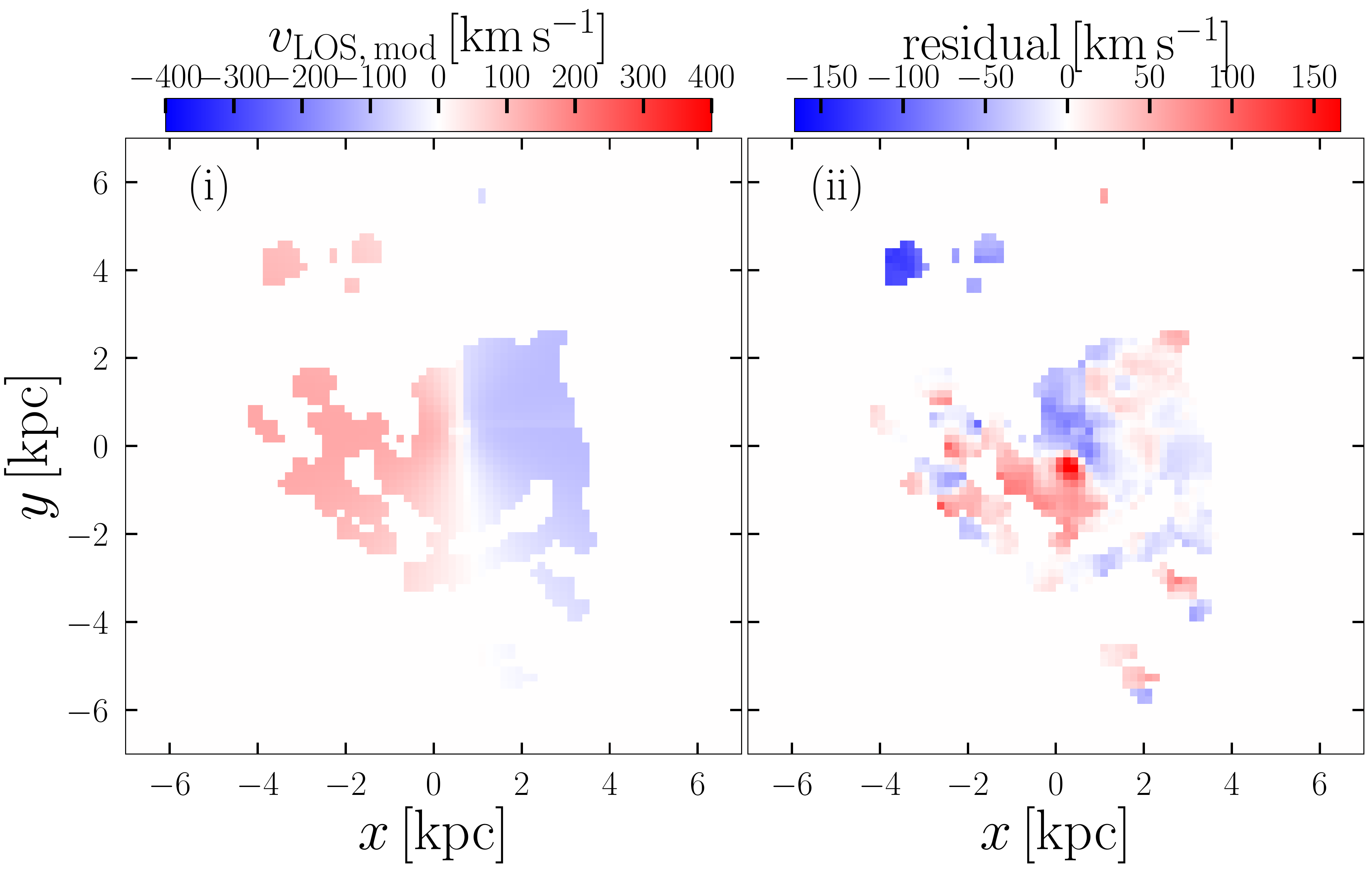} 
		\caption{Left: $v_{\rm LOS,mod}$ map (see equation Eq.\ref{eq:modvlos}) with model parameters of $20^{\circ}$, $0^{\circ}$, $15.6\kmsec$, $365.4\kmsec$ and 7.005 for $i,\,\theta_{0},\,v_{\rm sys}$, arctan asymptote and arctan gradient respectively. Right: map of residuals between the model and the $\vlosmi$ map  (panel \fpii\, of Fig.~\ref{fig:kinematics_3}), i.e. the former subtracted from the latter.
		}
	\label{fig:kinematics_4}
	\end{center}
\end{figure}

We present the kinematic structure of G1 post-\hsim observation in Fig.~\ref{fig:kinematics_3}. We again apply a SNR mask to remove contributions from spaxels with ${\rm SNR<3}$ before applying Eq.~\ref{eq:mom1} and \ref{eq:mom2}. From the $1^{\rm st}$ moment maps (panels \fpi\, and \fpii) we see that generally the $\vlos$ structure of G1 is preserved after observation with \hsimp. However there is a further reduction in detailed structure compared with the post-\lcars $\vlos$ maps.
 The $2^{\rm nd}$ moment maps are also fairly uniform, though there is a peak in $\sigmaobs$ at the galactic centre. Given how bright (see Fig.~\ref{fig:lcars_ha}) and chaotic the galactic centre is (see the lower panel Fig.~\ref{fig:vrot_actual}), seeing a high $\sigmaobs$ in this region is expected.

Passing the \lcars SSCs through \hsim introduces noise and instrumental effects. As a result a slit along the major axis of the galaxy cannot be used reliably to recover the intrinsic $v_{\theta}$. By adopting a simple kinematic model, it is possible to estimate the rotational velocity of G1 from \hsim observations. In this case we adopt Eq.~1 of \cite{Arribas:2008aa} in the form
\begin{equation}
	v_{\rm LOS,mod}(R,\theta) = \Omega(R)\cos{(\theta-\theta_{0})}\times\sin{i} + v_{\rm sys},
	\label{eq:modvlos}
\end{equation}
were $R$ is the distance from galactic centre, $\theta$ is the angle from the major axis, $v_{\rm LOS,mod}(R,\theta)$ is the line of sight velocity predicted by the model at $(R,\theta)$, $v_{\rm sys}$ is the systematic velocity of the galaxy along the line of sight, $i$ is the inclination angle of the galaxy, $\theta_{0}$ is the position angle and $\Omega(R)$ is a function describing the rotation curve (we again adopt an arctan profile). Fitting the above model to the $\vlosmi$ map shown in panel \fpii\, of Fig.~\ref{fig:kinematics_3}, setting $i=20^\circ,\,\theta_{0}=0^{\circ}$ and letting $\,v_{\rm sys}$, the asymptote value as well as the gradient of the arctan profile be unknowns, we find the model presented in Fig.~\ref{fig:kinematics_4}. In a general sense, the model is able to produce a crude approximation of $\vlosmi$, but with the bubbles of large $|\vlosmi|$ missing (as expected). From the simulation, it is known that the region with $\vlosmi\gtrsim200\kmsec$ centred around $(-1,-1)$ of panel \fpii\, of Fig.~\ref{fig:kinematics_3} is the result of a large number of young stars creating a low density, high velocity, high temperature bubble on the trailer side of the nearest spiral arm. The model presented in Eq.\ref{eq:modvlos} has no way to include, account for, or recreate stellar feedback driven gas in the model and thus is unable to produce a recreation of the \emph{turbulent} velocity structure of the galaxy. This leads to the large residuals seen in the right panel of Fig.~\ref{fig:kinematics_4}.

This model estimates the asymptote of the arctan function, i.e. $v_{\theta}$, to be $365.4 \pm5.5\kmsec$, which is slightly larger than the true value of $\sim340\kmsec$, but is consistent with the values found from the post-\lcars SSCs (when errors are considered). 
Given the inability to model the finer structure of a galaxy, we argue that such a model is not ideal for our all of our analysis, e.g. $\vsigma$ calculations see \S\ref{res:vsig}.

\begin{figure}
	\begin{center}
		\includegraphics[width=0.5\textwidth]{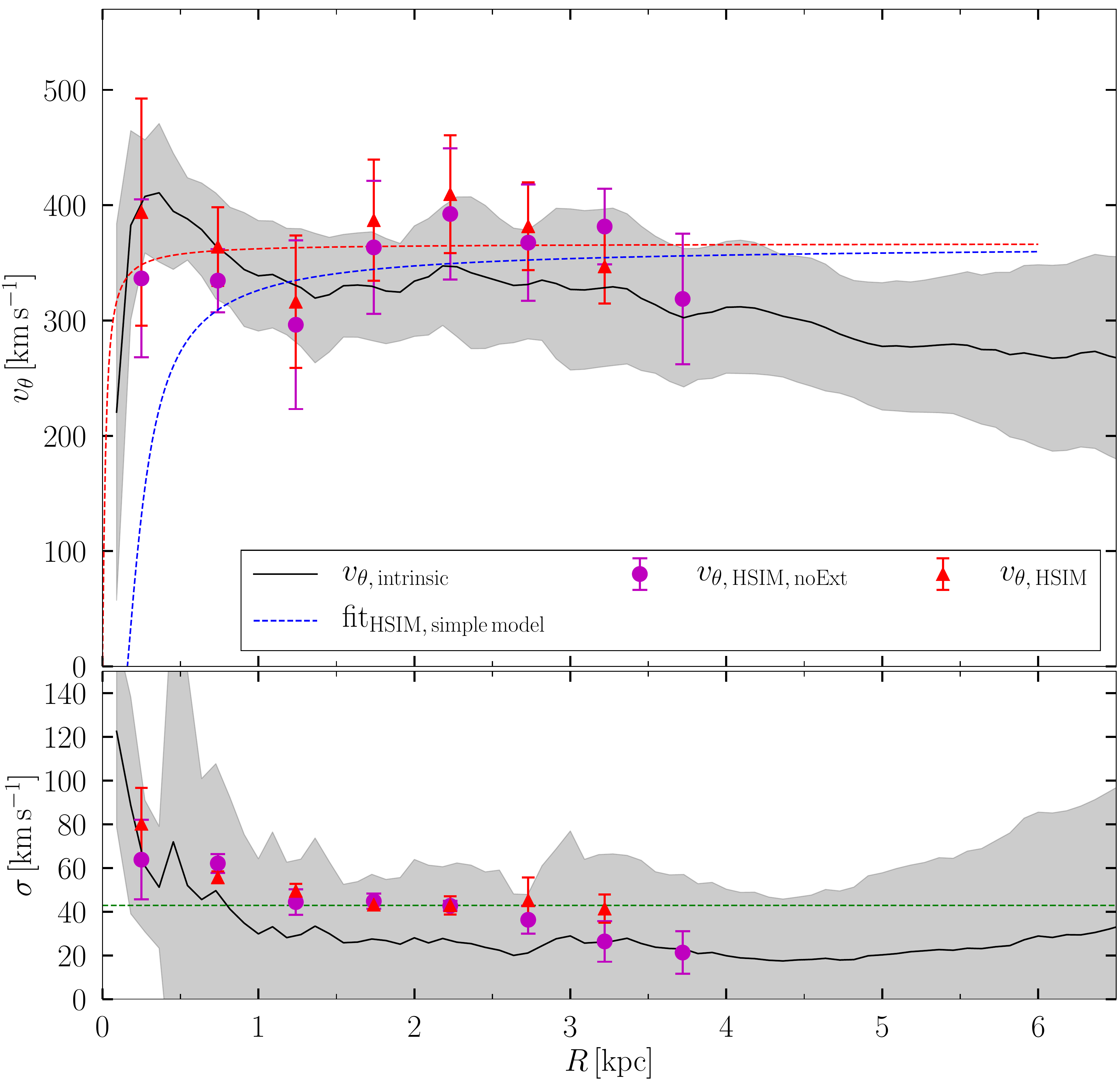} 
		\caption{Top: rotational velocity radial profile ($v_{\theta}(R)$). The solid black line shows the intrinsic curve for G1. The purple and red points show the values found by \barolo using the fiducial fitting parameters for G1 (no extinction and with extinction respectively) after being ``observed'' \hsimp. The red dashed line shows an arctan fit to the red (\barolo) data points, while the blue dashed line shows the arctan fit found by the simple kinematic model (i.e. Eq.~\ref{eq:modvlos}). The shaded region represents $1\sigma$ from the intrinsic curve. 		
		Bottom: velocity dispersion radial profile ($\sigma(R)$) for the same data sets as the top panel.
		}
	\label{fig:barolo_1}
	\end{center}
\end{figure}

We therefore opt to use a more sophisticated model that allows for clumps of gas at given $R$ to have high dispersion and $\vlosmi$. In this case we use the analysis tool \barolo \citep[see][for details]{Teodoro:2015aa} to extract a rotation curve from the map of $\vlosmi$. We provide \barolo the ``\texttt{reduced}'' and ``\texttt{reduced\_SNR}'' data cubes and allow the program to make its own SNR cuts to the data. In this case, it finds every pixel ${\rm SNR}>5$ as well as all the ${\rm SNR}>3$ pixels that are located around the first set of pixels. These regions are joined together if they are side by side. 
\barolo has a plethora of settings and variables that can be left to be determined or fixed. In this work we focus on 10 of these variables. We start by determining the best settings using our prior knowledge of the simulation and use this analysis run as our fiducial results from \barolo. To calculate errors on this fit we run \barolo seven additional times, for each of these runs we allow one or more variables to be determined by \barolo. The variables allowed to be fitted are: position angle, inclination angle, x-position of the galactic centre, y-position of the galactic centre, systematic velocity, radial velocity and position angle \emph{\&} inclination angle simualtanously. \barolo is able to weight the receding or approaching side of the galaxy preferentially or equally. In the fiducial settings we weight them equally.\footnote{When fitting the ``\texttt{reduced}'' cube produced from the SSC without extinction we preferentially fit to the approaching side of the galaxy as this provides the best match to the intrinsic rotation curve. This choice is discussed more in \S\ref{dis:weighting}.}
 By changing this, we are able to carry out 3 additional \barolo analysis runs: receding side preferentially weighted, approaching side preferentially weighted and position angle \& inclination angle with the approaching side preferentially weighted. To calculate an error on \barolop's fit we calculate the standard deviation of the ten non-fiducial runs from the fiducial run. The resulting rotation curve, with error is presented in Fig.~\ref{fig:barolo_1}.
 The fiducial \barolo model is in good agreement with the intrinsic curve. 
 
 As before we fit an arctan curve to the \barolo data points to determine the asymptote of the rotation curve and find a value of $366.9\pm17.3\kmsec$. 
 By repeating our mass estimate, i.e. $v_{\theta,\,{\rm BAROLO}}=v_{{\rm circ.Tot.}}=366.9\kmsec$, we estimate the total mass enclosed at $R=3.25$ to be $1.017\cdot10^{11}\pm10^{9.982}\Msol$ or $\sim0.81\times$ the total mass of G1. However, at $R=3.25$ the total mass enclosed in the simulation is $7.9\cdot10^{10}\Msol$ meaning that value calculated from observations is in fact an overestimate by factor of $\sim1.28$. As the estimated value is within $\sim\pm50\%$ of the actual value we argue that this is an excellent match and therefore that the total dynamical mass of a G1-like galaxy can be recoverable from observations with \hsim when combined with \barolo (or a \emph{similar}) during analysis. 
 
 It is worth noting that both values of $v_{\theta}$ calculated from the two methods used we have employed on the post-\hsim data cubes only differ by $\sim1.5\kmsec$, (which is an order of magnitude smaller than the error on the measured value of $v_{\theta}$). This would seem to indicate that either method will be reasonably reliable for actual HARMONI observations.

\subsection{$\vsigma$}
\label{res:vsig}

\begin{figure}
	\begin{center}
		\includegraphics[width=0.5\textwidth]{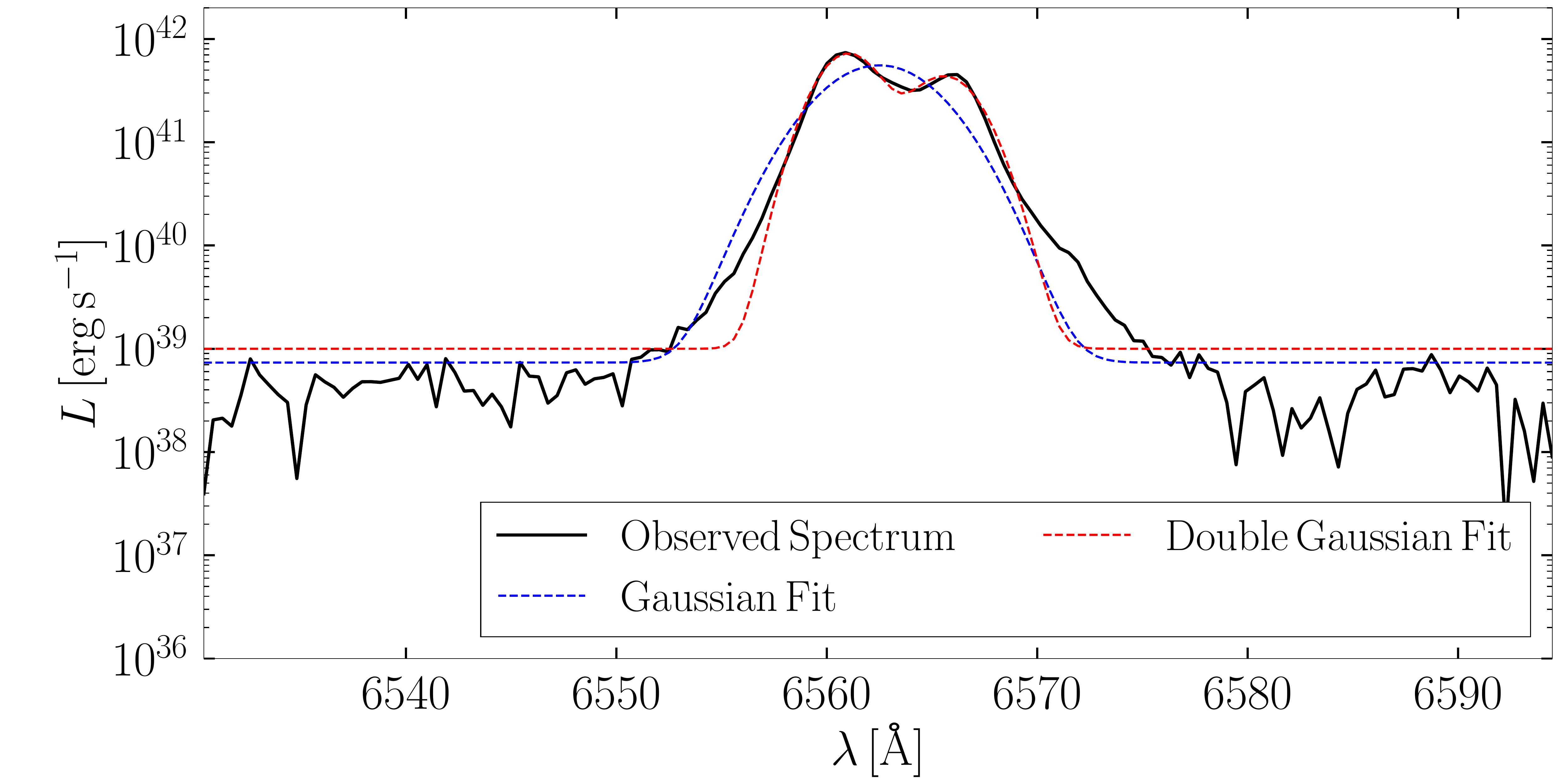} 
		\caption{Single aperture, continuum subtracted, spectrum of G1 after being observed with \hsim (black line). Dashed lines show a single Gaussian (blue) and double Gaussian (red) fit to the \ha emission line.
		}
	\label{fig:haemissionline}
	\end{center}
\end{figure}

\begin{figure*}
	\begin{center}
		\includegraphics[width=0.7\textwidth]{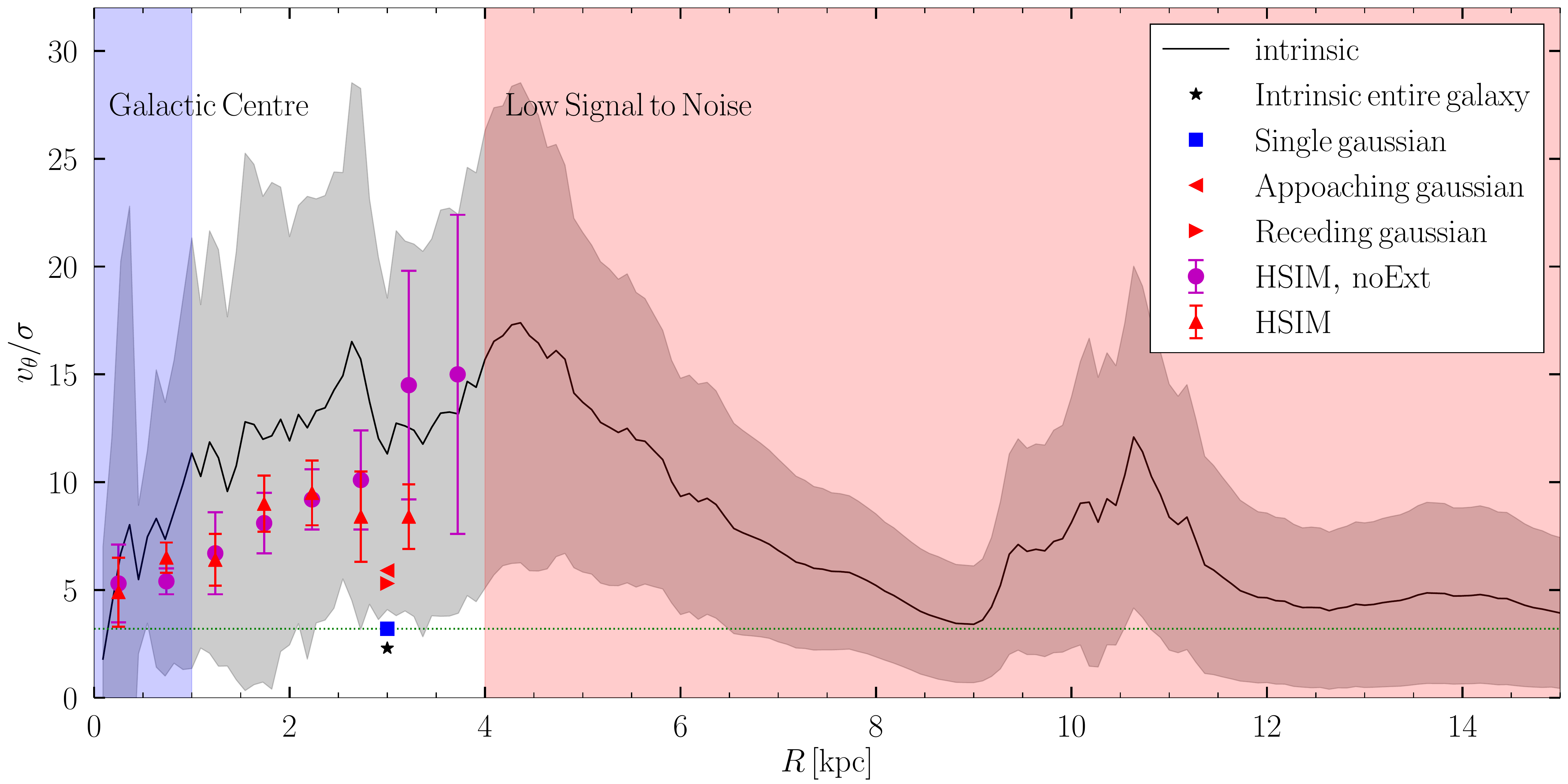} 
		\caption{Radial profile of $\vsigma$. The black line shows the intrinsic profile, with the coloured points showing values calculated using \barolo after G1 is observed by \hsimp. Red points show observations of SSC's which include extinction, while purple show those without extinction. The four data points with no error bars show the value of $\vsigma$ calculated on galaxy-wide scales (see \S\ref{res:vsig_galaxy_wide} for details). Those points do not depend on $R$ and are placed at $R=3\kpc$ by choice. The dotted green line shows $\vsigma=3.2$, i.e. the criteria for a merger.
		The grey shaded region shows the $1\sigma$ error on the intrinsic values. The blue shaded region indicates the G1's galactic centre, while shaded red region shows values of $R$ where the SNR is too low for \barolo to fit observations. 
		}
	\label{fig:vsigma}
	\end{center}
\end{figure*}

The dynamical ratio ($\vsigma$) has been used to characterise the dynamical state of observed galaxies for over 40 years \citep[][]{Binney:1978aa}. Galaxies that are isolated and rotationally supported against gravity have larger values of $\vsigma$ \citep[i.e. $>1$,][]{Johnson:2018aa} while those with smaller values are assumed to be unstable. \cite{{Pereira-Santaella:2019aa}} (henceforth PS19) adds further granularity by assuming those with $\vsigma<3.2$ are undergoing mergers.

\subsubsection{Measurements on Galaxy Scales}
\label{res:vsig_galaxy_wide}
Traditionally this calculation is carried out by calculating a value of $\sigma$ for the entire galaxy (normally from FWHM of an emission line of the galaxies integrated spectrum), while $v_{\theta}$ is taken from the flattest region of the rotation curve. By taking mass weighted mean of the galaxies intrinsic $\sigma$ values, we are able to calculate $\vsigma$ for G1 as $340\kmsec/150\kmsec\sim2.3$. Likewise using our arctan fit to the \barolo model and the fitting a Gaussian profile to the observed \ha emission line we find a value of $366.9\kmsec/114.2\kmsec\sim3.2$ post-\hsimp. The emission line for G1 is actually better fit by a double Gaussian than a simple Gaussian profile (see Fig.~\ref{fig:haemissionline}). This double Gaussian profile arises as a result of the receding side of the galaxy being distinguishable from the approaching side due to the spatial resolution of HARMONI. 
 Calculating $\vsigma$ for the broader Gaussian given $366.9\kmsec/68.8\kmsec\sim5.3$, while the slimmer gives $366.9\kmsec/61.9\kmsec\sim5.9$. All four of these $\vsigma$ suggest a gravitationally stable galaxy. If we adopt the PS19 criteria the first two calculated values would suggest that the galaxy is undergoing a merger while the last two suggest the opposite. 

Galaxies at high-$z$ are expected to be more turbulent due to a higher molecular gas fraction \citep[][]{Tacconi:2020aa}. The $\vsigma$ ratio for such galaxies would thus be lower and might be interpreted as a galaxy under going a merger when in fact it is still finding its equilibrium. To filter this effect, a redshift dependant, threshold velocity dispersion ($\sigma_{0}$) can be applied \citep[see PS19 and][for a more complete discussion]{Ubler:2019aa}. PS19 advocates that to classify a galaxy as isolated and stable requires $\sigma<\sigma_{0}$ \emph{and} $\vsigma>3.2$. Different previous works have defined value of $\sigma_{0}$ in different ways. In this work we use Eq.~2 of \cite{Ubler:2019aa} (as present in Eq.~5 of H21) to calculate the ``typical'' value of $\sigma$ at $z=2.2$ as our threshold. Thus we calculate $\sigma_{0}=42.9\pm6.0$ at $z=2.2$. Combining the $\sigma_{0}$ and $\vsigma$ criteria to the observed value and it is not clear whether G1 would be classed as undergoing a merger if measured on a galaxy wide scale, due to both a high $\vsigma$ and $\sigma>\sigma_{0}$. If the double Gaussian is resolved and used, either side would suggest a stable/isolated galaxy. These results demonstrate that measuring $\vsigma$ on galaxy wide scales does not provide a clear cut indicator of a settled disc or merger galaxy as normally thought. We discuss the interpretation of $\vsigma$ further in \S\ref{dis:vsigma}.

\subsubsection{Measurements on Sub-Kiloparsec Scales}
\label{res:vsig_small_scales}

As outlined above (see \S\ref{lcars:spec}), it is possible to calculate $\sigmasim$ for each cell in the simulation which in turn allows for maps and radial profiles of $\sigmasim$ to be calculated (see Fig.~\ref{fig:kinematics_1} and \ref{fig:vrot_actual} respectively). We use the intrinsic radial profiles of $v_{\theta}$ and $\sigmasim$ to calculate the intrinsic radial profile of $\vsigma$, which is presented in Fig.~\ref{fig:vsigma}. From visual inspections of G1 we know that the galaxy is a disc with spiral arms (see Fig.~\ref{fig:smaps}) and therefore would expect $\vsigma>1$, which we find at all radii. With the exception of $R<0.5\kpc$ we find $\vsigma>5$ for G1. As noted previously (see \S\ref{sect:galaxydiscrip}) the centre of the galaxy tends to have high values of $\sigmasim$ and hence a reduced $\vsigma$.
At larger radii, particularly those capturing the connective gas between G1 and G2, $\vsigma$ decreases however even here the lowest $\vsigma$ reaches is $3.4$, which according to the classifications in PS19 would be interpreted as an isolated galaxy. It is worth noting that $\vsigma$ increases within G2, indicating that the orbiting velocity of G2 about G1 dominates over the signal of any internal $\sigma$ or $v_{\theta}$ of G2. Taking the mean of all $\vsigma$ values within G1 gives $\langle\vsigma\rangle=12.3\pm9.1$, again confirming the classification of rotationally supported disc. It is worth noting that that $\langle\vsigma\rangle$ does not match the value of $\vsigma$ calculated in \S\ref{res:vsig_galaxy_wide}, the implications of this discrepancy are discussed  in \S\ref{dis:vsigma}

Unlike our simple kinematic model (see \S\ref{hsim:kinematic}) \barolo calculates a $\sigma$ value for each ring that it calculates $v_{\theta}$ for. The $\sigma$ radial profile determined by \barolo is a good match to intrinsic profile, with high $\sigma$ near the galactic centre and smaller values at larger radii (see lower panel of Fig.~\ref{fig:barolo_1}). 
Using the analysis from \barolo we are able to calculate the observed radial profile of $\vsigma$ (red data points on Fig.~\ref{fig:vsigma}). While there are differences between the intrinsic and observed radial profile of $\vsigma$ in general they are in good agreement. For example, within the central kiloparsec $\vsigma$ hovers around $5$ while for $1<R<3.5\kpc$ $\vsigma$ sits between $5$ and $10$. As before, due to the SNR cut and low SNR values at $R>3.5$ \barolo is unable to provide any information of $\vsigma$ at large radii. For $1<R<4\kpc$ we find $\langle\vsigma\rangle=7.6\pm1.4$ ($\langle\vsigma\rangle=9.3\pm2.8$ when extinction is neglected). As with the intrinsic measurement this value of $\langle\vsigma\rangle$ would normally be interpreted as G1 being a rotationally supported disc galaxy.  
\begin{figure*}
	\begin{center}
		\includegraphics[width=0.9\textwidth]{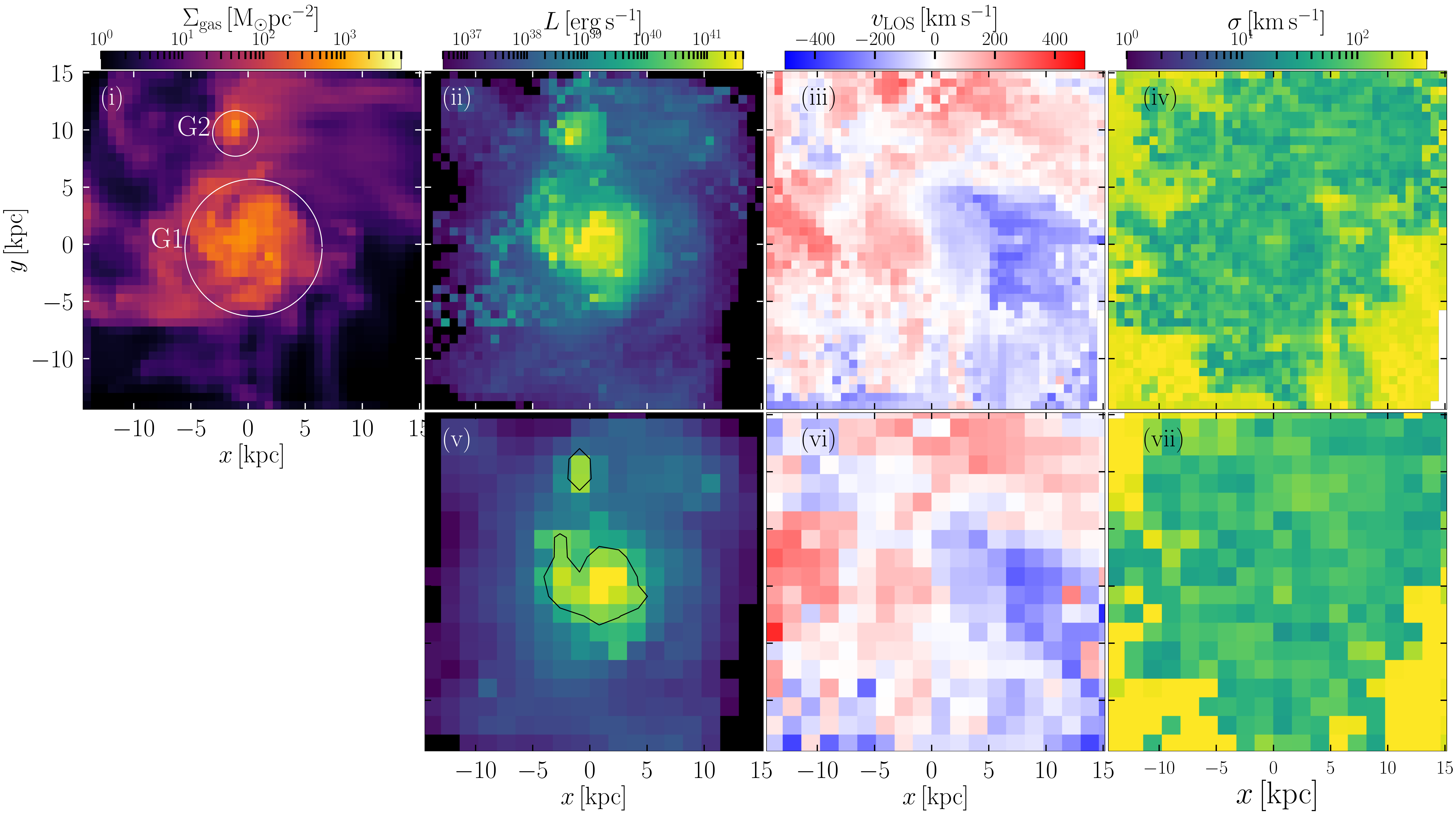} 
		\caption{Impact of Resolution on intensity maps. Top row: maps of G1 with a spatial resolution of $\Delta x=722\pc$. Panel \fpi\, shows the gas surface density map. The white circles are identical to the circles in Fig.~\ref{fig:smaps}. Panels \fpii\, \fpiii\, and \fpiv\, show the corresponding \ha $0^{\rm th}$, $1^{\rm st}$ and $2^{\rm nd}$ moment maps once G1 is processed by \lcarsp. Bottom row: Mock KMOS observations of the fiducial \lcars data cube (see text for details).  Panels \fpv\, \fpvi\, and \fpvii\, show the \ha $0^{\rm th}$, $1^{\rm st}$ and $2^{\rm nd}$ moment maps respectively. The contour shown on \fpv\, represents the pixels likely to be detected after applying a SNR cut.
		}
	\label{fig:lowres}
	\end{center}
\end{figure*}
For G1 the intrinsic velocity dispersion is less than $\sigma_{0}$ with the exception of $R<1\kpc$ (i.e. galactic centre) and $R>8\kpc$ (i.e. outside of G1), where we find values as large as $\sim123\kmsec$ (see lower panel of Fig.~\ref{fig:barolo_1}). Similarly we find that the values of $\sigma$ determined by \barolo are very close to $\sigma_{0}$ for $R>1.5$. From both of the above two quantitive metrics G1 would be classified as an isolated, rotationally supported, stable galaxy. Yet G1 is known to be at the second periapsis of a merger with G2 (see Fig.~\ref{fig:sfh}) at the time of our analysis. The lack of a merger signature in $\vsigma$ and $\sigma$ could simply be a result G2 being significantly smaller (only $12.5\%$ of the gas mass) than G1 or the impact parameter of the merger. However the simplest explanation is that G2 is still too far away from G1 at the time of analysis. $\sigma$ and hence $\vsigma$ may only show merger signatures once the G2 crosses through the gas or stellar discs. In Grisdale et al., (in prep) we explore this signature and impact of mergers on G1.

\section{Discussion}
\label{sect:discusion}

\subsection{Impact of Resolution}
\label{dis:resolution}

\begin{figure*}
	\begin{center}
		\includegraphics[width=0.65\textwidth]{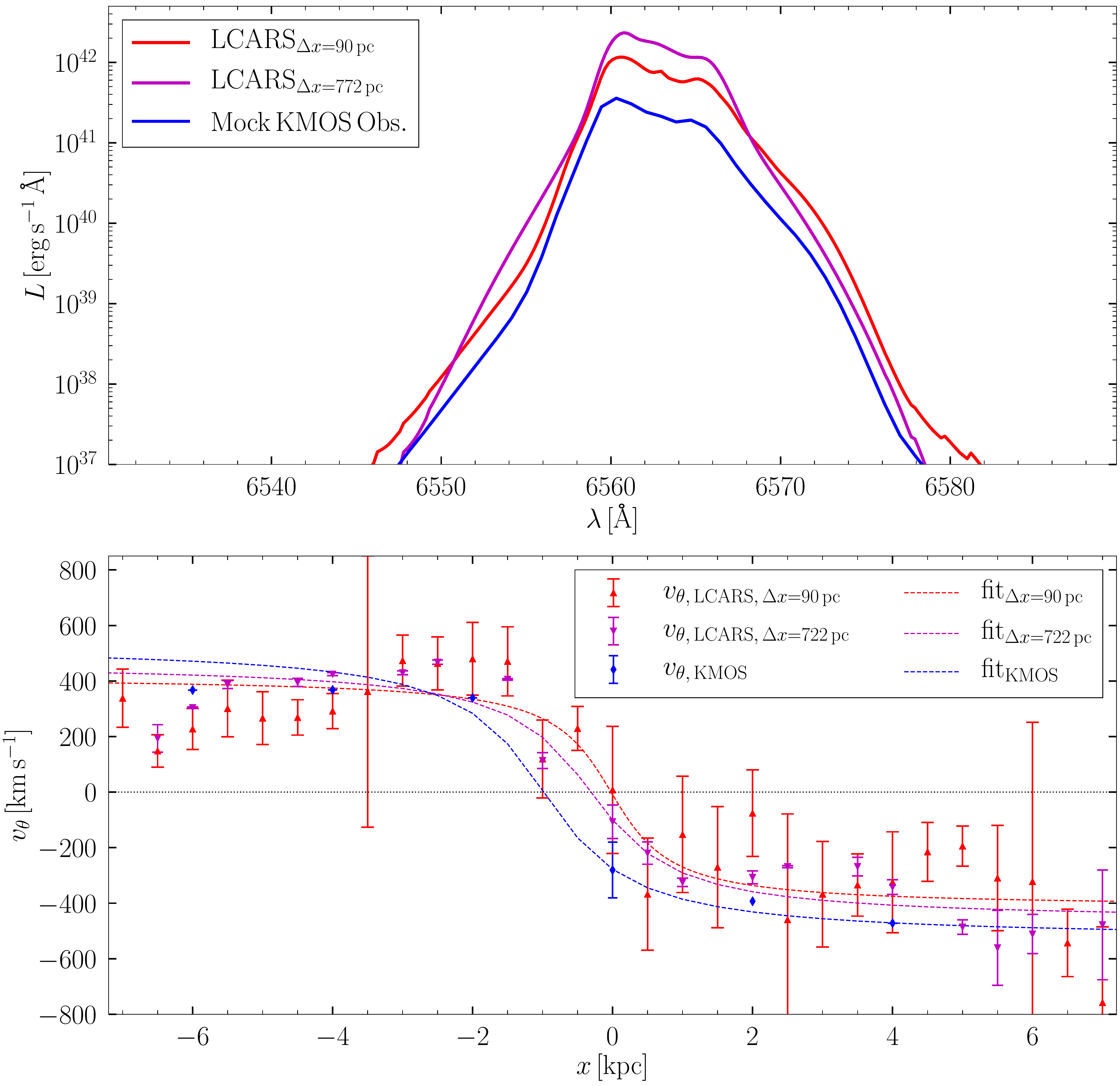} 
		\caption{Top: Comparison of the impact that different resolutions have on the single aperture (continuum subtracted) spectrum of G1. The red, purple and blue lines show: the fiducial resolution, \lcars run at to $\Delta x=722\pc$ and a mock KMOS observation of the fiducial \lcars outputs respectively. 
		Bottom: Comparison of the rotational velocities when calculated from a slit for the three different resolutions. Dashed lines show arctan fits to the corresponding data points.}
	\label{fig:lowresspec}
	\end{center}
\end{figure*}

\subsubsection{Resolution of \lcars}
\label{dis:resolution1}
All the results in the previous sections have been calculated using a fixed spatial resolution of $\Delta x\sim90\pc$ for data cube passed to \lcars and then \hsim. To test the impact of the choice of spatial resolution, we rerun \lcars on G1 with a spatial resolution $8$ times lower, i.e. $\Delta x=722\pc$. In Fig.~\ref{fig:lowres} we present the pre-\lcars surface density map, as well as the \ha $0^{\rm th}$, $1^{\rm st}$ and $2^{\rm nd}$ moment maps post-\lcars. At $z=2.2$ this corresponds to an angular pixel size of $\sim86\mas$ which is larger than the biggest pixel size available in \hsim ($60\mas$), for this reason we do not pass these data cubes through \hsimp, instead we carry out our analysis on the \lcars SSC. At this lower resolution the G1 and G2 are still distinguishable in both the surface density map and the post-\lcars \ha map. However in both cases G2 has lost the definition of its internal structure. Some internal structure is visible in G1 however without prior knowledge of it would be next to impossible to determine what form that structure takes. 

By comparing the single aperture spectrum of the $\Delta x=722\pc$ run to fiducial run, we see that the dual gaussian profile remains (see the top panel of Fig.~\ref{fig:lowresspec}) though the receding peak is reduced and the approaching peak is enhanced. Changing the resolution of \lcars does not change the number of star particles in the simulation but does reduce the number of SSPs and increases the number of stars that contribute to each SSP. Therefore the same number of photons are concentrated into a smaller number of cells, particularly closer to the galactic centre. This results in the more concentrated luminosity seen in Fig.~\ref{fig:lowresspec}. 

Turning to the kinematics, there is again a significant reduction in structures of both the $1^{\rm st}$ an $2^{\rm nd}$ moment maps. With that being said, comparing panels \fpiii\, and \fpiv\, of Fig.~\ref{fig:lowres} with panels \fpii\, and \fpiv\, of Fig.~\ref{fig:kinematics_1} shows a general agreement with the kinematic maps created at the fiducial resolution. By applying the same slit based approach (see \S\ref{res:lcars:kinem}) to the $1^{\rm st}$ moment map of the $\Delta x=722\pc$ run we are able to produce a rotation curve for this \lcars run, see the lower panel of Fig.~\ref{fig:lowresspec}. From which we estimate $v_{\theta}\sim467\kmsec$, which would equate to a total mass of $10^{11.48}\Msol$ or $\sim2.4\times$ the actual total mass of G1. Allowing $x_{\theta}$ to be a fitted parameter we see that reducing the resolution moves $x_{\theta}$ to $\sim-310\pc$ which is $\sim785\pc$ from the true value. When $x_{\theta}$ is fitted for in the fiducial resolution run we found $x_{\theta}\sim-22\pc$. In either case the difference is on the order of a single cell and reasonably close to the actual galactic centre.

\subsubsection{Resolution of Observations: KMOS vs. HARMONI}
\label{dis:resolution2}
In reality it will be the distance to a galaxy, as well as the sensitivity and resolution power of the instrumentation used to observe it that will determine the resolution of a data cube used for analysis. Testing this can be achieved by taking the output SSCs from \lcarsp, reducing the spatial resolution and spectral resolution as well as applying a Point Spread Function (PSF). For comparisons with H21 we opt to reduce the spatial and spectral resolutions to $0.19''$ ($1.62\kpc$) and $2.8\ang$ respectively. Additionally we apply a Gaussian PSF with a FWHM of $0.6''$ to mimic values of H21's observations with the KMOS on the VLT.\footnote{This is very simple model for a mock observation and should only be used as guide on how observations with HARMONI will differ to those on KMOS.} As one might expect, this mock KMOS observation, presented in panel \fpv\, of Fig.~\ref{fig:lowres}, shows a significantly degraded \ha $0^{\rm th}$ moment map. The sites of star formation can no longer be seen and all traces of spiral arms have been lost. Calculating a single spectrum from the mock KMOS observations produces a similar, double gaussian \ha emission line (see top panel of Fig.~\ref{fig:lowresspec}). Due to the much broader PSF of the mock observation, some photons from the emission line have become indistinguishable from the continuum, which results in the luminosity of the emission being reduced to $\sim0.3$ of full resolution \lcars \ha emission line. This effect is normally accounted for using a point source with a known luminosity to calibrate the observations.

Panels \fpvi\, and \fpvii\, of Fig.~\ref{fig:lowres} shows the $1^{\rm st}$ and $2^{\rm nd}$ moment maps for these mock observations. At the resolution of these mock observations the kinematic structure of the galaxy is, again, severely degraded. We again apply a slit across the $1^{\rm st}$ moment map and calculate a rotation curved (shown in the lower panel of Fig.~\ref{fig:lowresspec}). This calculation estimates $v_{\theta}=533\kmsec$ and $x_{\theta}=-513\pc$. This places the calculated centre of rotation almost a kiloparsec from the intrinsic centre of rotation, and overestimates the G1's rotation speed by a factor of $\sim1.6\times$ faster that its true value. Given the very low spatial resolution and the wide PSF, this significant overestimate is understandable and highlights the importance of being able to resolve a galaxy sufficiently that kinematic measurements are representative of the observed galaxy. 

Until this point we have assumed that mock KMOS observations have a ${\rm SNR}>3$ for all spaxels. Realistically this is not the case and low SNR spaxels should be accounted for in our mock observations. To this end we assume that the highest signal to noise found is $20$ and that this is found in the most luminous pixel of the $0^{\rm th}$ moment map. We mask all pixels that fall below one twentieth of the luminosity of that pixel (pixel above this cut are shown via the contour in panel \fpv\, of Fig.~\ref{fig:lowres}). With this SNR cut in place G1 loses further structural definition and G2 is reduced to a single pixel. The measured \sfr\, of the galaxy is negligibly affected by the SNR cut. There is an insufficient number of pixels remaining in the $1^{\rm st}$ moment map to calculate a trustable $v_{\theta}$ value and for this reason we do not attempt to.
In short once the prospect of noise is accounted for it is no longer possible to recover, let alone realistic, kinematics from our mock KMOS observations.

We assert that a physical comparison between panel \fpiii\, of Fig.~\ref{fig:hsim_msk} and panel \fpv\, of Fig.~\ref{fig:lowres} presents a strong argument for the need for high spatial and spectral resolution offered by the $30\,{\rm m}$-class telescope. Furthermore, without relying on a proxy, such as $\vsigma$, detection of minor mergers is extremely difficult, as is the detection of inflows and outflows. As discussed in \S\ref{dis:vsigma}, even $\vsigma$ has issues associated with it when calculated on large scales, and so should be used with caution. Simply put observations on the \eelt in general and with HARMONI specifically have the potential to be game changing in the exploration of galaxies at high redshifts ($z\gtrsim2$). 

It is worth noting that the \eelt will allow galaxies at a range of redshifts to be probed on scales of $\sim100\pc$. If these new observations of high redshift galaxies have clumpy structures, similar to simulated galaxy in this work, new analysis tools will likely be required. We will be exploring this in upcoming work.

\subsection{Are \ha Photons a Good Tracers for Gas Kinematic Properties?}
\label{dis:weighting}

This work focuses on exploring whether the intrinsic properties of a galaxy can be recovered from observations, such as those taken with HARMONI on the \Eelt. Our results show that in general this is the case, however there are noticeable differences between intrinsic and observed properties. For example, as noted in \S\ref{res:lcars:kinem} and seen in Fig.~\ref{fig:kinematics_1}, the map $\vlosmi$ has several pockets of gas moving at speeds on the order of $500\kmsec$ which are not present in the map of $\vlosma$. To properly understand how the observations link back to intrinsic properties of a galaxy it is necessary to understand where these high velocity regions come from.  

The answer to this question is weighting. In the case of the intrinsic map each pixel in the map is a mass weighted average of all cells along the line of sight, while in the $\vlosmi$ map each pixel is the result from the  photons summed along the line of sight. When mass weighting, the spiral arms dominate the rotation signature at a given $R$. However in the latter case photons are removed due to extinction in the arms. Photons that have to travel through dense gas, such as the spiral arms, have their contribution to $\vlosmi$ reduced. Areas of the galaxy with low density gas (e.g. inter-arm gas) are affected to a lesser degree and so their contributions are effectively boosted. 

We see an example of this is the region found at $(-3.0,0.5)$ on panel \fpi\, of Fig.~\ref{fig:smaps} which corresponds to the $\vlosmi\sim500\kmsec$ region found at the same location in panel \fpii\, of Fig.~\ref{fig:kinematics_1}. This region is located behind a spiral arm and ahead of two large gas clouds as a result has very little extinction meaning that the brightest cells in this region will dominate the measurement of $\vlosmi$. 
From panels \fpii\,and \fpiii\, of Fig.~\ref{fig:smaps} we know there  are only ``old'' stars in this region, i.e. all with ages greater than $10\Myr$. Such stars are unlikely to provide a sufficient injection of energy to drive gas to such velocities. Instead we have to look to the surrounding gas, which does have young stars. Panel \fpiii\,of Fig.~\ref{fig:kinematics_1} shows this region to have mass weighted $\sigma$ in excess of $30\kmsec$, while panel \fpiv\, shows values of $>200\kmsec$. Combining all the available information we are able to determine the source of such high velocity regions seen only in the photon ``weighted'' maps. The galactic disc contains regions of low density, inter-arm gas. This is being heated by young stars in the surrounding arms and gas clumps, which accelerates the gas to speeds of several hundred kilometres a second. The old stars plus diffuse photons from the neighbouring young stars (see \S\ref{lcars:diffuse}) illuminate the region sufficiently for the high $\vlosmi$ to be seen in panel \fpii\, of Fig.~\ref{fig:kinematics_1}. However, when mass weighting to create the $\vlosma$ map the gas ``above'' and ``below'' the galaxy contribute to $\vlosma$ and bring the mean value down. 
By taking a slice through G1 at the correct depth it is possible to created $\vlos$ map that will show the high velocity region seen at $(-3.0,0.5)$. By creating slices at multiple depths it would be possible map all the high velocity pocket within G1. 

This effect is still present after the SSC has been passed through \hsim and is at least partially responsible for the larger value of $v_{\theta}$ found at $R=2.25\kpc$. Indeed, by looking at \fpii\, of Fig.~\ref{fig:kinematics_3} there are small regions or ``bubbles'' of high $v_{\rm LOS,\,obs}$ on the approaching side of the galaxy. Weighting one side of the galaxy preferentially over the other when carrying out analysis can reduce $v_{\theta}$ in several places, by ignoring such regions. Its also worth noting that we found that stellar feedback driven outflows and corresponding inflows (as gas falls back on to the galaxy) could impact the values of $v_{\theta}$ and $\sigma$ determined by \barolo. Again these are low density, high velocity structures and thus have low mass but strong photon weighting. This result highlights how a small scale structures within the galaxy can impact the measured physical properties recovered from observations.

Despite the discussion above it is clear that the spiral arms and galactic centre still provide the majority of the \ha emissions in our \hsim observations (see Fig.~\ref{fig:lcars_ha} and \ref{fig:hsim_msk}). Furthermore, as we showed throughout \S\ref{sect:resultsi} it is possible to recover reasonable measurements of G1's $\sfr$, rotation curve and dynamical ratio profile. In summary, there is limited fungibility between mass and photon weighting when determining the properties of a galaxy however it is possible to use the latter to get good estimates of a galaxies intrinsic properties. In the case of all real galaxies in the physical Universe it is highly unlikely we will be able to directly measure their properties as we can in simulations, therefore it is highly important to know the limits of how observations can be translated back to the galaxy's physical properties.

\subsection{Importance of $\vsigma$} 
\label{dis:vsigma}

Section~\ref{res:vsig} showed various ways of calculating the dynamical ratio, $\vsigma$, with each method resulting in different calculated values. All of the methods used to calculate $\vsigma$ in this work have found G1 to be a rotationally supported disc, however the method used can change the measured value by as much as a factor $\sim6$. If this ratio is to be used as a measure of a galaxies gravitational stability, or an indicator of a merger, this presents a problem: which method should be considered, which is valid and when is it valid? We discuss those difficulties here. 

The first method looked at the global properties of a galaxy (e.g. width of single aperture spectrum and the flattest part of the rotation curve, see \S\ref{res:vsig_galaxy_wide}), in this work $\vsigma$ was calculated on scales of $12\kpc$ (i.e. the diameter of G1). This kind of approach makes sense when the internal structure of a galaxy can't be resolved, such as when current generation of $10{\,\rm m}$-class telescope carry out observations of galaxies at $z\sim2$, e.g. see our mock KMOS observations above or Fig.~2 of H21.
 Due to the $\sim4.75\times$ larger increase in the primary mirror diameter combined with the adaptive optics of \eelt we see a $\sim9.8\times$ increase in spatial resolution at $z\sim2.2$ with the \eelt over the VLT. This allows for structures such as spiral arms to be resolved within the galaxy as well as being able to resolve the velocity dispersion on sub-kiloparsec scales and thus radial profiles of $v_{\theta}$, $\sigma$ \& $\vsigma$ to be calculated. 

In \S\ref{dis:weighting} we discussed the impact of using photon weighting rather than using mass weighting when calculating properties of galaxies. Here we see another example of the impact the type of weighting has. When taking the single aperture spectrum of a galaxy the brightest regions will dominate the spectrum, furthermore bright regions with low $\sigma$ will contribute more to the shape of the emission line than equally bright (i.e. the integral of the emission line is the same), high $\sigma$ regions as the latter can become lost in the continuum. These complications need to be considered when using a single aperture spectrum to determine $\sigma$ and $\vsigma$. 

We have already seen that the expected spatial resolution of observations with HARMONI should allow for $\sigma$ to be resolved on sub-kiloparsec scales and thus for a determination of how $\vsigma$ changes within a galaxy (e.g. see Fig.~\ref{fig:vsigma}) becomes possible. As a result we need to (re)evaluate what $\vsigma$ on such scales mean. The intrinsic $\vsigma$ profile is mass weighted as a result will be dominated by the velocity structures of arms (see \S\ref{dis:weighting}). Therefore the $\vsigma$ profile provides a measure of whether internal velocity structures of clouds and spiral arms at a given radius are more important than galactic rotation.   
For example, if at some radius $\vsigma$ is found to be much greater than $1$, it would be a strong indicator that GMCs or spiral arms have relatively little internal turbulence and are simply clumps of gas moving around within the galaxy. However if $\vsigma$ is $\leq1$ for a given $R$, this would suggest that the GMCs are highly turbulent, to such a degree that their internal motions are comparable to the speed at which the cloud is orbiting the galaxy. Finally, if $\vsigma<<1$ is found at a given $R$, this would be a strong indicator that gas structures are either rapidly collapsing due to gravity or being blow apart (most likely by stellar feedback).

Such interpretations are complicated when inferred from observations, where brightest regions \emph{may} not be tracing mass. That being said, given our discussion in \S\ref{dis:weighting} it is clear that the \ha emission lines in our mock observations at least partially trace the gas mass. In which case the above interpretation of a galaxies radial $\vsigma$ profile should hold. 

Calculating the mean dynamical ratio for the galaxy from the radial profiles ($\langle\vsigma\rangle$) also proved to be different to galaxy-scale calculation. The brightest part of G1 is the galaxy's centre (see Fig.~\ref{fig:hsim_msk}), as a result this region has a significant impact on the shape of the \ha emission line for galaxy. Especially given that the highest values of $\sigma$ are found at $R<1$, i.e. the galactic centre. Furthermore, while $v_{\theta}$ tends to be higher in this region, it is only by a factor of $\sim1.2$, compared with $\sigma$ increasing by a factor of $\gtrsim2.5$ and thus $\vsigma$ in the galactic centre is lower. In essence the traditional observer's method of measuring a single value of $\sigma$ and $v_{\theta}$ for the galaxy is biased towards the galaxies centre where $\sigma$ is higher, while $\langle\vsigma\rangle$ givens a measurement that account for the entire galaxy. With the above in mind we would not expect to see signature of a minor merger in single spectrum calculated $\vsigma$ until either 1) the two galaxies are close enough to disrupt the galactic centre of the primary galaxy or 2) there is sufficient star formation occurring, in the gas connecting the two galaxies, that the galactic centre no longer dominates the single-aperture spectrum. 

Finally, we highlight that all of our calculations of $\vsigma$ are carried out at specific scales, e.g. galaxy wide,  $500\pc$ rings, etc. As shown in \cite{Agertz:2015aa} the scale used when measuring any stability criteria will impact the apparent stability of a galaxy. We therefore remind the reader that with any criteria used to determine stability, even a simple one such as $\vsigma$ it is important to consider which scales the calculation is carried out on.

\section{Conclusion}
\label{sect:conclusion}
In this work we explored whether it is possible to accurately recover the intrinsic properties of a galaxy at a redshift of $z\sim2.2$ with the \eelt and the HARMONI spectrograph. To achieve this we passed a galaxy, from the cosmological simulations \newhorizonp, through the post-processing pipeline \lcars (v2.0) to add photons to the simulations. The resulting \ha Spatial-Spectral Cube (SSC) were then observed using the HARMONI simulator (\hsimp) and analysed. The galaxy used in this work, which we label G1, is the primary galaxy in a merging pair. At the time of this analysis the secondary galaxy, G2, is at the periapsis of its second orbit of G1. Our key results are as follows:

\begin{enumerate}
	\item After being processed with \lcars and observed with \hsim the morphology of the G1 remains visible: spiral arms are clearly seen, as is a bright galactic centre and the more defuse gaseous disc. Post-\hsim the morphology more closely resembles a map of the young (ages $<10\Myr$) stars within the galaxy. 
	
	\item Measuring the \sfr\, before and after observation recovers values that are larger than the intrinsic \sfr\, of the galaxy ($38.3\Msolyr$). In the case of the former we recover a value of $65.8\Msolyr$ when correcting for extinction and $59.1\Msolyr$ if extinction is not added by \lcarsp. For the latter we find $\sfr=69.3\Msolyr$ when correcting for extinction. Both sets of values are within a factor of two of the intrinsic value.
		
	\item $1^{\rm st}$moments maps of G1 show that the line of sight velocity structure of G1 is largely recoverable from the \ha emission line. After observation there is a reduction in the detail of the $1^{\rm st}$moments map however the key features remain. The $2^{\rm nd}$moment map of G1's SSC shows larger regions of high velocity dispersion ($\sigma$) than the intrinsic map. These high $\sigma$ are reduced in the observed $2^{\rm nd}$moment map due to originating in the inter-arm, low density, non-star forming gas, i.e. regions with low SNR.
	
	\item Applying a slit to the  $1^{\rm st}$moments map, perpendicular to the the axis of inclination, allow for G1's rotation curve to be recovered with good accuracy from the SSCs. After being observed too much internal structure is lost for a slit to be effective. An analysis tool, such as \barolo, is able to produce a rotation curve that reasonably matches the intrinsic curve of the galaxy. From these curves an estimate of the G1's mass is recoverable and found to be within $\pm40\%$ of the intrinsic total mass. 
	
	\item We calculate the dynamical ratio $(\vsigma)$ in several ways. In all cases finding G1 to be considered rotationally support against gravity (i.e. $\vsigma>1$). Calculating $\vsigma$ on a galaxy wide scale (using the peak of the rotational curve and width of a single aperture spectrum), finds a value of $2.3$ intrinsically and $3.2$ after observation. However both of these values are dominated by the galaxies centre and are not representative of the entire galaxy. Calculating $\vsigma$ as a function of radius using the galaxies rotation curve and radial $\sigma$ profile shows that the majority of the galaxy has $\vsigma>10$ ($>5$ from \hsim observations). Furthermore, averaging the $\vsigma$ profile provides a very different global picture of the galaxy the compared with $\vsigma$ calculated on a galaxy wide scale, e.g. $7.6$ vs. $3.2$, as the former uses data from throughout the galaxy while the latter is dominated to the galactic centre. We also note that the galactic scale calculation can be influenced by the interpretation of the emission line, e.g. is it a single or double gaussian profile and thus what is its width. 
\end{enumerate}
	
While we have shown that the physical/intrinsic properties of a galaxy can be recovered to a reasonable degree from photons emitted by that galaxy, we have also shown that emissions can be impacted by low density gas and thus do not provide an accurate representation of the properties of the majority of the gas along a line of sight. In short, measuring properties via emission lines is not a replacement for being able to measure the gas properties directly. However, given that the latter is likely impossible the former will have to be sufficient.

In conclusion, by using a HARMONI on the \eelt it will be possible to calculate the physical properties of a galaxy at $z\sim2.2$, including its \sfr, kinematic structure, rotation curve, total mass, and $\vsigma$ profile. However, we advise careful consideration of methods as how low SNR regions are handled can impact the \sfr\, measurement, while choice of scale can lead to drastically different interpretations of how much a galaxy is rotationally supported. Finally, we emphasise that the primary goal of this work is to determine if the galaxy's intrinsic properties are recoverable. To that end the settings used for the \hsim observations and analysis are tuned to provide the best possible result.  
 
\section*{acknowledgments}
We thank the referee for the useful and constructive comments. 
KG and NT acknowledge support from the Science and Technology Facilities Council (grants ST/N002717/1 and ST/S001409/1), as part of the UK E-ELT Programme at the University of Oxford.
KG, LH, DR, JD, AS and IGB acknowledge support from STFC through grant ST/S000488/1.
The research of JD and AS is supported by Adrian Beecroft and STFC.
MPS acknowledges support from the Comunidad de Madrid through Atracci\'{o}n de Talento Investigador Grant 2018-T1/TIC-11035 and and PID2019-105423GA-I00 (MCIU/AEI/FEDER,UE). 
SKY acknowledges support from the Korean National Research Foundation (NRF-2020R1A2C3003769). 
KK acknowledges support from the DEEPDIP project (ANR-19-CE31-0023).
This project has received funding from the European Research Council (ERC) under the European Union?s Horizon 2020 research and innovation programme (grant agreement No 693024)
This work was granted access to the HPC resources of CINES under the allocations c2016047637, A0020407637 and A0070402192 by Genci, KSC-2017-G2-0003 by KISTI, and as a ``Grand Challeng'' project granted by GENCI on the AMD Rome extension of the Joliot Curie supercomputer at TGCC. 
The supercomputing time for numerical simulation was kindly provided by KISTI (KSC-2017-G2-003), and large data transfer was supported by KREONET, which is managed and operated by KISTI. 
Calculations were performed with version 17.01 of \cloudy \citep{Ferland:2017aa} and \hsim version 3.03 (see \url{https://github.com/HARMONI-ELT/HSIM}).
We thank JWB for fruitful discussions. 

\section*{Data Availability}
The data underlying this article will be shared on reasonable request to the corresponding author.

\bibliographystyle{mn3e}
\bibliography{ref_local}

\end{document}